\documentclass[11pt]{article}

 \font\gotb eufm10 scaled \magstep1
 \newcommand{\df}{{\sc Definition}}
\newcommand{\rem}{{\sc Remark}}
\newcommand{\post}{{\sc Postulate}}
\newcommand{\exm}{{\sc Examples:}}
\newcommand{\st}{{\sc Proposition}}
\newcommand{\bb}{\bibitem}
\newcommand{\cc}{\cite}
\newcommand{\vp}{\varphi}
\newcommand{\vt}{\vartheta}
\newcommand{\sss}{\sigma}
\newcommand{\al}{\alpha}
\newcommand{\Om}{\Omega}
\newcommand{\om}{\omega}
\newcommand{\lt}{\left}
\newcommand{\rt}{\right}
\newcommand{\lll}{\lambda}
\newcommand{\F}{{\cal F}}
\newcommand{\OO}{{\cal O}}
\newcommand{\D}{\hat D}
\newcommand{\am}{\hat a^-}
\newcommand{\ap}{\hat a^+}
\newcommand{\U}{\hat U}
\newcommand{\V}{\hat V}
\newcommand{\W}{\hat W}

\newcommand{\s}{\hat S}
\newcommand{\K}{\hat K}
\newcommand{\I}{\hat I}
\newcommand{\E}{\hat E}
\newcommand{\Aa}{\tilde A}
\newcommand{\A}{\hat A}
\newcommand{\B}{\hat B}
\newcommand{\p}{\hat p}
\newcommand{\X}{\hat X}

\newcommand{\po}{\Psi_0}
\newcommand{\aA}{F_A}
\newcommand{\HH}{\hat H}
\newcommand{\AAA}{\hbox{\gotb A}}
\newcommand{\LL}{\hbox{\gotb L}}
\newcommand{\QQ}{\hbox{\gotb Q}}
\newcommand{\xxi}{\hbox{\boldmath$\xi$}}
\newcommand{\ttau}{\hbox{\boldmath$\tau$}}
\newcommand {\QQQ}{\hbox {\gotb Q}_{\xi}}
\newcommand {\qqq}{\hbox {\gotb Q}_{\xi'}}
\newcommand {\vx}{\vp_{\xi}}
\newcommand {\vpx}{\vx(\A)}
\newcommand{\BB}{\hbox{\gotb B}}
\newcommand{\JJ}{\hbox{\gotb J}}
\newcommand{\HHH}{\hbox{\gotb H}}
\newcommand{\bea}{\begin{eqnarray} \label}
\newcommand{\eeq}{\end{equation}}
\newcommand{\beq}{\begin{equation} \label}
\newcommand{\eea}{\end{eqnarray}}
\newcommand{\nn}{\\ \nonumber}
\newcommand{\rr}[1]{(\ref{#1})}

 \author{D.A.Slavnov}

\title{Measurements and Mathematical Formalism of Quantum Mechanics
\thanks{Physics of Particles and Nuclei, 2007, Vol. 38, No. 2, pp. 147-176.} }
   \date{}

\begin{document}

  \maketitle
    \begin{center} {\it  Department of Physics, Moscow State
University,\\  Moscow 119899, Russia. E-mail:
slavnov@goa.bog.msu.ru }
 \end{center}

 \begin{abstract}

A scheme for constructing quantum mechanics is given that does not
have Hilbert space and linear operators as its basic elements.
Instead, a version of algebraic approach is considered. Elements
of a noncommutative algebra (observables) and functionals on this
algebra (elementary states) associated with results of single
measurements are used as primary components of the scheme. On the
one hand, it is possible to use within the scheme the formalism of
the standard (Kolmogorov) probability theory, and, on the other
hand, it is possible to reproduce the mathematical formalism of
standard quantum mechanics, and to study the limits of its
applicability. A short outline is given of the necessary material
from the theory of algebras and probability theory. It is
described how the mathematical scheme of the paper agrees with the
theory of quantum measurements, and avoids quantum paradoxes.
\end{abstract}

\section {INTRODUCTION}

The development of quantum theory has revolutionized physics. This
may be a cliche, but it is the truth. Indeed, within quantum
physics, a huge number of phenomena have been described that
resisted treatment within classical physics. A plethora of new
technologies have been developed based on quantum physics.

However, like any revolution, the quantum revolution has its back
side. Imperceptively, substitution of notions had taken place in
physics. In quantum physics, "to explain a phenomenon" means "to
give a mathematical description of the phenomenon."

The origin of this substitution is understandable. Modern quantum
physics is an axiomatic theory based on mathematical
axioms~\cc{von}. The axioms are very convenient for constructing a
powerful mathematical formalism. At the same time, the axioms are
almost completely detached from our intuitive notions~\cc{seg}.
Most theoreticians in physics support the opinion that physical
intuition based on classical concepts is useless in quantum
theory. Thus, theory can be constructed starting from a more or
less arbitrary set of mathematical axioms. The only requirements
are that the axioms should be consistent, and the consequences
they imply should account for a wide enough range of experimental
data. In this way, in passing, the quantum revolution substituted
explanations of physical phenomena with mathematical descriptions.

Modern quantum mechanics is based on the following postulates:

$I.$ The state of a physical system is described with a vector
$|\Psi\rangle$ of a Hilbert space, or with a statistical operator
(density matrix) in this space;

$II.$ Observables  $\D$ of the system are described with self
adjoint operators  $\hat {\cal D}$;

$III.$ The average value of observable  $\D$ over the state
$|\Psi\rangle$ equals the expectation value $\langle\Psi|
\hat{\cal D}|\Psi\rangle$.

Why does Hilbert space have any relation to the state of a
physical system? Why does operator $\hat {\cal D}$ correspond to
the observable $\D$? Why is the average value of the observable
$\langle \Psi|\hat {\cal D}|\Psi\rangle$. All these questions are
considered impertinent.

The slogan "Victors need never explain, you cannot question
success" has overcome standard quantum mechanics. The huge number
of excellent results obtained from the above axioms allows one to
wave away the impertinent whys with a clear conscience.

Despite this, doubts remain. On the other hand, the excellent
results obtained with the above axioms cannot be attributed to
coincidences. May it be that statements  $I-III$ should not be
taken as primary postulates, that they would better be associated
with more fundamental theses related more directly to physics?

If so, it should be possible to expose conditions for validity of
statements  $I-III$. In other words, it should be possible to
establish the range of applicability of quantum mechanics. This
may settle the controversies over the quantum paradoxes, which
have agitated the physics community starting almost from the
origin of quantum mechanics.

They break relations between quantum and classical physics. In the
latter, the states and the observables are described with
completely different mathematical constructs. Overall, the
classical-to-quantum relation comes out rather strange. On the one
hand, classical physics is considered to be a limiting case of
quantum physics, i.e., it is a derivable theory. On the other
hand, formulation of quantum mechanics requires the notion of
interaction of a quantum object with a measuring device obeying
the classical description~\cc{land}. Logically, this is a vicious
circle. To break it, one states that classical logic is not valid
in quantum physics, and it is to be replaced with quantum logic.

In this way, quantum theory implies, on top of other things, a
revolution in logic. However, in contrast to the quantum
revolution in physics, the revolution in logic has not been
productive. Besides, a consistent and self contained quantum logic
is yet to be created. The isolated statements available on this
subject are either a reformulation of the postulates  $I-III$, or
their corollary. Consequently, in physical practice, statements of
quantum logic are not used, and the underlying postulates $I-III$
are used directly.

There is another annoyance in the standard formulation of quantum
mechanics. In its basics, quantum mechanics is a statistical
theory. Therefore, it should be based on probability theory.
Presently, probability theory (in the Kolmogorov
formulation~\cc{kol})is a rather solidified mathematical subject.
However, it is assumed that this probability theory is not fit for
quantum mechanics, and a dedicated quantum probability theory is
required. Thus, on top of the above, quantum theory requires a
revolution in mathematics. As with quantum logic, no success has
been achieved in this direction. At best, detached statements are
available for the new probability theory, and, in fact, they are
again corollaries of the postulates $I-III$ (e.g., see \cc{hol}).

We conclude that there is a gap between quantum theory and
mathematics; the latter stays within classical logic and
probability theory. The above implies that it is highly desirable
to design a mathematical scheme that would be equally fit for both
classical and quantum physics. It would be nice, if the rules of
the game, or, in philosophers' parlance, the paradigm of the
scheme were classical. By classical paradigm, we mean here one
that respects classical formal logic and incorporates the notion
of causative relations between physical phenomena, as well as
between logical statements. Next, it should imply the existence of
physical realities that carry causes of physical phenomena.
Additionally, it would conjecture that probability propositions
obey classical Kolmogorov probability theory.

Usually, it is assumed that the above statements are incompatible
with the mathematical scheme accepted in quantum mechanics. Here,
we attempt to prove the opposite. Our exposition is not based on
postulates  $I-III$. Instead, we use the algebraic
approach~\cc{emch,hor,blot}. We formulate axioms within this
approach that are, firstly, more fundamental than the postulates
$I-III$, and, secondly, are much more amenable to intuitive
understanding~\cc{slav1,slav2,slav3}.

Admittedly, a psychological barrier must be crossed at this point.
The formalism of Hilbert space has become standard in quantum
mechanics. Due to this, it is perceived as intuitively
understandable. Physical intuition is replaced here with a
particular mathematical formalism. In contrast, the formalism of
algebraic theory is not so familiar to most physicists. Due to
this, statements employing the language developed in the theory of
algebras are perceived as more involved than the parallel
statements employing the language of Hilbert space. This is
despite the fact that the algebraic statements are more elementary
as a rule. We help to step over this psychological barrier in the
next section with an outline of elementary facts taken from the
theory of algebras.

\section {ELEMENTS OF THE THEORY OF ALGEBRAS}

We outsource definitions and statements
from~\cc{emch,rud,naj,brat,dix}.

\

\df{} 1. A set  \LL{} is called a complex (real) linear space if\\
 (a)  for any complex (real) number  $\al$ and any element
  $\U \in \LL$, there is a unique element $\al \U \in \LL $;\\
(b)  for any two elements $\U,\V \in \LL$, there is a unique
element  $\U+\V \in \LL$;\\
(c)  operations (a) and (b) have the
familiar properties of multiplication and addition, respectively.

\

\df{} 2. A complex (real) linear space $\LL${} is called a complex
(real) algebra \AAA{} if a multiplication operation is defined for
any elements $\U,\V,\W \in \AAA$ that satisfy the following
properties:\\
  (a) $\U\V\in\AAA$;\\
  (b) $(\U+\V)\W=\U\W+\V\W$, $\U(\V+\W)=\U\V+\U\W$;\\
  (c) $\al(\U\V)=(\al\U)\V=\U(\al\V)$.

  \

\df{} 3. Algebra  \AAA{} is called an associative algebra if the
relation $\U(\V\W)=(\U\V)\W$ holds for any elements
$\U,\V,\W\in\AAA$.

\

\df{} 4. Algebra \AAA{} is called a commutative algebra if the
relation $\U\V=\V\U$ holds for any elements $\U,\V\in\AAA$.

\

\exm\\
 (a) the set of all real continuous bounded functions of one
variable is a real algebra;\\
 (b) the set of all complex continuous bounded functions is a complex
  algebra;\\
 (c)  the set of bounded linear operators of a Hilbert space is a complex
algebra;\\
 (d) the set of mutually commuting bounded Hermitian
linear operators of a Hilbert space is a real algebra;\\
 (e) the set of all bounded Hermitian linear operators of a Hilbert space
is not an algebra.

   \

\df{} 5. A mapping  $\U\to\U^*$ of a complex algebra \AAA{} onto
itself (i.e., $(\U,\U^*\in\AAA)$) is called an involution if the
following holds for any complex number $\al$ and any
$\U,\V\in\AAA$:\\
 (a) $(\U+\V)^*=\U^*+\V^*$,\\
 (b) $(\al\U)^*=\al^*\U^*$,\\
 (c) $(\U\V)^*=\V^*\U^*$,\\
 (d) $\U^{**}=\U$.

 \

 \exm\\
(a) a complex conjugation is an involution for \AAA{}that is the
set of all complex bounded functions of a single variable;\\
 (b) a Hermitian conjugation is an involution for \AAA{} that is the
set of all bounded linear operators of a Hilbert space.

 \

\df{} 6. A complex algebra equipped with an involution operation
is called an involutive algebra.

 \

\rem{}. The identity transformation is an involution for any real commutative algebra.

 \

\df{} 7. If $\U^*=\U$ ($\U\in\AAA$), the element  $\U$ is called
Hermitian.

 \

\df{} 8. An element  $\I\in\AAA$ satisfying the relations
$\I\U=\U\I=\U$ for any  $\U\in\AAA$ is called the unit element of
the algebra.

 \

\st{} 1. Any algebra either has a unit element or can be
supplemented with an element satisfying the properties of the unit
element.

 \

 In the following, we consider the algebras with unity.

 \

 \df{} 9. Element $\U^{-1}\in\AAA$ is called the inverse of
 $\U$, if $\U^{-1}\U=\U\U^{-1}=\I.$

\

\df{} 10. The spectrum  $\sss(\U;\AAA)$ of the element  $\U$ in
the algebra  \AAA{} ($\U\in\AAA$) is the set of all numbers $\lll$
for which the element  $\lll\I-\U$ has no inverse element within
the algebra \AAA.

\

\df{} 11. The number $r=\sup\{|\lll|; \quad
\lll\in\sss(\U;\AAA)\}$ is called the spectral radius of the
element $\U$.

\

\df{} 12. A subset  $\QQ$ of an algebra  \AAA{} is called a
subalgebra if  $\QQ$ is an algebra for the available definition of
multiplication and addition.

\

\df{} 13. Let $\QQ$ be a real commutative subalgebra of algebra
\AAA. The subalgebra  $\QQ$ is called a maximal commutative real
subalgebra if it is not contained in any other such subalgebra of
\AAA.

\

Generally, the spectrum  $\sss(\U;\QQ)$ of element  $\U$ with
respect to algebra  \QQ{} may not coincide with the spectr
$\sss(\U;\AAA)$ of the same element with respect to algebra  \AAA.
However, the following proposition holds.

\

\st{} 2. If \QQ{} is a maximal real commutative subalgebra of
algebra  \AAA{} and $\U\in\QQ$, then $\sss(\U;\QQ)=\sss(\U;\AAA)$.

\

\df{} 14. A set  $\JJ_l$ of elements belonging to algebra  \AAA{}
is called its left ideal if:\\
 (a) $\JJ_l\neq\AAA$,\\
 (b) $\JJ_l$ is a linear subspace of  \AAA,\\
 (c) relations $\U\in \JJ_l$, $\V\in\AAA$ imply  $\V\U\in \JJ_l$.

 \

Right ideals are defined similarly. A set  $\JJ$ of elements from
algebra \AAA that is simultaneously a left and right ideal is
called two-sided ideal.

\

\df{} 15. Let $\JJ$ be a two-sided ideal of algebra \AAA. Elements
$\U$ and $\V$ are called equivalent with respect to $\JJ$ if
$\U-\V\in \JJ$. A set of all elements equivalent to each other is
called a residue class of algebra \AAA.

\

\df{} 16. The set of all residue classes of algebra  \AAA{} is
called a factor algebra, and denoted  $\AAA/\JJ$.

\

\st{} 3. The set  $\AAA/\JJ$ equipped with operations of class
multiplication by numbers and class additions introduced as
respective operations for representatives of the classes becomes
an algebra. In other words, a factor-algebra is an algebra.

\

\df{} 17. An  involutive  algebra  is  called normed if a norm
$\|\U\|$ is defined for each element $\U$. The norm is a
nonnegative number satisfying the following conditions:\\
 (a) $\|\al\U\|=|\al|\|\U\|$;\\
 (b) $\|\U+\V\|\leq\|\U\|+\|\V\|$;\\
 (c) $\|\U^*\|=\|\U\|$;\\
 (d) $\|\U\V\|\leq\|\U\|\;\|\V\|$;\\
 (e) $\|\U\|=0$ implies $\U=0$.

 \

\df{} 18. A quantity  $\|\U\|$ satisfying all the above conditions
apart of condition (e) is called a semi norm.

 \

 \df{} 19. A sequence of elements of a normed
spac $\{\U_n\}$ is called fundamental if for any $\varepsilon>0$
there exists a number  $N(\varepsilon)$, such that the
inequalities $n>N(\varepsilon)$ and $m>N(\varepsilon)$ imply the
inequality  $\|\U_n-\U_m\|<\varepsilon$.

 \

\df{} 20. A normed space in which any fundamental sequence is
convergent in norm to an element of this space is called complete.
 \

\df{} 21. A complete normed space is called a Banach space.

 \

\st{} 4. Any normed space can be completed to a Banach space.

 \

\df{} 22. An involutive associative algebra that is a Banach space
(a Banach algebra) with the norm satisfying the additional
requirement  $\|\U^*\U\|=\|\U\|^2$ is called a C*-algebra.

 \

\df{} 23. A mapping $\U\to\U'$ of an involutive algebra  \AAA{}
 $(\U\in\AAA)$ into involutive algebra  $\AAA'$ $(\U'\in\AAA')$
is called a homomorphism from algebra  \AAA{} to algebra $\AAA'$
if $\U\to\U'$ and $\V\to\V'$ implies the following relations:
$\U^*\to\U^{'*}$, $\al\U\to\al\U'$,
 $\U+\V\to\U'+\V'$, and $\U\V\to\U'\V'$.

 \

A homomorphism may map several elements of  \AAA{} into a single
element of  $\AAA'$.

 \

\df{} 24. If a homomorphism is a one-to-one mapping, it is called
an isomorphism.

 \

\df{} 25. An isomorphic mapping of an algebra onto itself is
called an automorphism.

 \

\df{} 26. A homomorphism of a commutative associative real
(complex) algebra \AAA{} into the set of real (complex) numbers is
called a character of the algebra.

 \

\df{} 27. A homomorphism of algebra  \AAA{} into a set of linear
operators of a Hilbert space \HHH{} is called a representation of
the algebra.

 \

\df{} 28. A mapping from normed algebra  \AAA{} to normed algebra
$\AAA'$ is called isometric if  $\U\to\U'$ implies
$\|\U\|\to\|\U'\|$.

 \

\df{} 29. Mapping $\U\to\vp(\U)$ from algebra  \AAA{} into complex
numbers is called a linear functional if
 $\vp(\al\U)=\al\vp(\U)$ and $\vp(\U+\V)=\vp(\U)+\vp(\V)$. Here,
 $\U,\V\in\AAA$ and $\al$, $\vp(\U)$ are complex numbers.

 \

\df{} 30. Linear functional  $\vp$ on an involutive algebra
 \AAA{} is called positive if  $\vp(\U\U^*)\geq 0$
 for any  $\U\in\AAA$.

 \

 \st{} 5. If $\vp(\U)$ is a positive functional, then\\
  (a) $\vp(\U^*)=\vp^*(\U)$,\\
  (b) $|\vp(\U^*\V)|^2\leq\vp(\U^*\U)\vp(\V^*\V)$.

  \

\st{} 6. A positive functional on a Banach algebra is continuous.

  \

\st{} 7. If $\vp(\U)$ $(\U\in\AAA)$ is a character of an
associative commutative algebra  \AAA, then\\
  (a) $\vp(0)=0$,\\
  (b) $\vp(\I)=1$,\\
  (c) $\vp(\U\U^*)\geq 0$.

  \

Thus, character is a positive functional on algebra  \AAA.

  \

\st{} 8. For a Banach algebra \AAA{}, the following holds for the
set $\{\vp(\U)\}$ of all its characters:\\
  (a) $\lll=\vp(\U)\in\sss(\U;\AAA)$;\\
  (b) if $\lll\in\sss(\U;\AAA)$, there exists $\vp(\U)\in\{\vp(\U)\}$
  for which  $\lll=\vp(\U)$.

  \

\df{} 31. Element $\p$ of algebra  \AAA{} is called a projector if
$\p^*=\p$, $\p^2=\p$.

  \

\df{} 32. Projector  $\p_{\lll}\neq 0$ is called minimal if
relations $\p_{\lll}\p_{\mu}=\p_{\mu}\p_{\lll}=\p_{\mu}$imply
either $\p_{\mu}=0$ or $\p_{\mu}=\p_{\lll}$.

  \

\st{} 9. If \AAA{} is an algebra of bounded linear operators in a
Hilbert space  $\HHH$, any minimal projector is a projector onto a
one-dimensional subspace of space  $\HHH$.

  \

\df{} 33. Sequence $\{\U_n\}$ of elements of algebra  \AAA{} is
called convergent in weak topology to an element $\U$ if the
relation $\vp(\U_n)\to \vp(\U)$ holds for any linear bounded
positive functional $\vp$.

\

\df{} 34. A set $G$ of elements belonging to Banach algebra \AAA{}
is called a generating set of the algebra if the smallest closed
subalgebra containing  $G$ coincides with  \AAA.

  \

\df{} 35. The Boolean algebra of a set  $\Om$ is the totality of
all subsets of  $\Om$ supplied with the following operations:\\
(a) operation of logical addition --- the union of subsets;\\
 (b) operation of logical multiplication --- the intersection of subsets;\\
 (c)  operation of logical negation --- the complement of subsets to the
 set $\Om$.

  \

\df{} 36. A Boolean algebra is called complete with respect to an
algebraic operation if the operation results in an element of the
original algebra.

  \

\df{} 37. A Boolean algebra is called $\sss$-algebra under the
following conditions:\\ (a) it contains the set $\Om$ and the
empty set $\emptyset$;\\
 (b) it contains the complement to $\Om$
of any subset belonging to the algebra;\\
 (c) it is closed under
countable number of unions and intersections of subsets.

   \

\df{} 38. A set  $\Om$ with a specific $\sss$-algebra selected is
called a measurable set.

   \

Hereafter, we refer the above definitions and propositions in the
following style: (D.35.b) means definition 35 point b, (P.7.c)
means proposition 7, point ñ.

\section {OBSERVABLES, MEASUREMENTS, AND STATES}

Let us turn to physics. We aim at formulating the basic postulates
of quantum mechanics. We will try to use as postulates statements
admitting direct experimental checks, in contrast to checks of
distant consequences. Otherwise, we risk making redundant
assumptions and running into contradictions. Correspondingly, we
start with physical phenomena, and tune the mathematical
formalism, instead of looking for a physical interpretation of a
preset mathematical scheme.

The basic notion in studying physical systems is the notion of the
observable. It seems to be a self evident notion not requiring an
exact definition. Heuristically, an observable is an attribute of
a physical system that can be supplied with a number value via a
measurement procedure.

\

\rem. In what follows, we assume that units are fixed, and all the
observables can be considered as dimensionless.

\

Both classical and quantum physics have observables as their basic
notions. Despite this, in mathematical formalisms of classical and
quantum physics, the observables correspond to different
mathematical objects. Let us attempt unification. Aiming at this,
we try to isolate the relevant mathematical features of
observables, separating them form the ones that are customarily
attributed to observables for purposes of facilitating development
of the mathematical formalism.

\

\rem. Not infrequently, a number of observables are not subject to
changes for a system under study. For example, in the study of
interactions between photons and electrons, the masses of
electrons and photons and the charge of an electron can be given
beforehand. It is convenient to exclude such quantities from the
set of observables, considering them as parameters involved in the
definition of the system.

\

Under a measurement, the system is influenced by the measuring
device. Depending on the kind of this influence, measurements can
be classified into two kinds: reproducible and irreproducible.
Reproducible measurements are characterized by the property that,
despite the perturbation introduced upon each measurement,
repeated measurement of the same observable with the same or some
other device yields the same result. It is assumed that between
the measurements, the system is not affected by external
influences, and we are able to take into account the changes in
the values of the observables due to free evolution.

The problem of reproducibility is of particular interest in the
case when a number of observables are measured for the same
physical system. Let us first measure observabl $\A$, then
observable  $\B$, then again observable  $\A$ (probably, with a
different device), and, finally, observable $\B$. If the results
of the repeated measurements coincide with the original ones, we
call such measurements compatible. If there are devices allowing
one to make combined measurements of observables $\A$ and $\B$, we
call such observables compatible or simultaneously measurable.

Experiments demonstrate that all the observables are compatible
for classical systems. In contrast to this case, there are both
compatible and incompatible observables in the quantum case.

In the standard quantum mechanics, this fact is incorporated into
the complementarity principle~\cc{bohr1}. We simply consider it as
a manifestation of the necessity of incompatible devices for
measuring two incompatible observables~\cc{zeil}.

We denote the set of all observables by  $\AAA_+$, and its maximal
subset of compatible observables by  $\QQQ$. The subscript  $\xi$
distinguishes between different maximal subsets of incompatible
observables. The subscript  $\xi$ takes values in the set  $\Xi$.
Evidently,  $\Xi$ consists of a single element for a classical
system. For a quantum system, this set contains more than one
element. Subsequently, we will see that this set is infinite, and
even has the cardinality of continuum. One and the same observable
may simultaneously belong to different subsets  $\QQQ$.

Experiments demonstrate that for any two compatible observables
$\A$ and $\B$, there exists a third observable $\D$ possessing the
following properties. First, it is compatible both with  $\A$ and
$\B$. Second, the outcomes of simultaneous measurements of the
observables $\A$, $\B$, and $\D$ (for one and the same physical
system) satisfy the relation

  \beq{1}
 A+B=D.
\eeq

In fact, simultaneity is not particularly relevant. Compatibility
of these observables suffices. Despite this, for brevity, we
characterize such a situation saying that the observables are
measured simultaneously.

Relation~\rr{1} holds regardless of the particular outcomes of the
measurements. In view of this, it is possible to assume that the
observables themselves are related by a similar relation:

$$\A+\B=\D.$$

In this way, it is possible to equip a set  $\QQQ$ with addition.
Analogously, multiplication of elements, and multiplication by
real numbers are introduced. Experiments show that each of the
subsets  $\QQQ$ has the properties of real associative commutative
algebra. Thus, the mathematical description of observables
maintains that they are elements of some algebra. Until now, we
have justified this statement only for compatible observables. We
will see that it can be extended to incompatible observables as
well.

Relating to each physical observable (via compatible measurements)
the result of a measurement,

$$\A\to A=\vx(\A),$$ we define a functional on the algebra $\QQQ$.
By the definition of algebraic operations in  $\QQQ$, this
functional is one of the characters of algebra $\QQQ$ (see
(D.26)).

Any real measurement of an observable gives it a finite value.
Reversing this fact, physical observables (i.e., the ones measured
in real experiments) are only the observables $\A$ satisfying the
relation
  \beq{4}
 \sup_{\xi}\,\sup_{\vx}|\vx(\A)|<\infty.
 \eeq

In the following, we will see that the boundedness of functionals
$\vx(\cdot)$ is not an insurmountable obstacle for considering
within the theory unbounded observables. Such observables are
commonplace in standard quantum mechanics.

We summarize the above considerations in the following postulates

\

\post{} 1.  A set  $\QQQ$ of compatible observables can be
equipped with the structure of a real associative commutative
algebra. Conversely, if observables belong to one and the same
real associative commutative algebra, they are compatible.

\

\post{} 2. For a classical system, all observables are compatible.

\

\post{} 3. The outcomes of simultaneous measurements of
observables belonging to algebra  $\QQQ$ are described by real
bounded (in the meaning of inequality~\rr{4}) functional
$\vx(\cdot)$, which is a character of algebra $\QQQ$.

\

Above, we repeatedly used the notion of physical system. It is
intuitively clear what this notion means. In view of this, we will
not try to give it a physical interpretation. However,
constructing mathematical formalism requires clear understanding
of the meaning of the expression "a given physical system."

In the following, we assume that a physical system is given if a
number of conditions are met. First, a set of system observables
$\AAA_+$ is given. Quantities whose values are known beforehand
and are not varied are not included in $\AAA_+$ Instead, their
values are considered given. Second, sets of compatible
observables  $\QQQ$ $(\xi\in\Xi)$. are given. These sets are
subsets of set  $\AAA_+$. Third, relations between observables are
given. Relations between compatible observables are defined by
their membership in  $\QQQ$. Relations between incompatible
observables will be considered later. Fourth, the dynamics of the
system is given. Since we aim at unified description of classical
and quantum systems, we will not use a concrete method of
describing system dynamics. We simply take that observables can
depend on time, and the statement "relations between observables
are given" implies also that relations between observables at
consecutive moments of time are given.

\

\rem. If the physical system under consideration is a conservative
one, because of uniformity of time, it is appropriate to take that
dynamics is described with a time-dependent automorphism on the
algebra of observables. However, if the system endures external
perturbations depending on time, description of dynamics may be
more involved. In particular, it may require modifications of the
algebra of observables itself.

\

Apart from consideration of the whole Universe as a given physical
system, any physical system is a subsystem of a larger system.
Mathematically, this means that the observables of the system form
a subset of observables of another system. Selection of the subset
can be carried out by various features. The first among these is
localization.

Any domain  $\OO$ of the four-dimensional space-time is related to
a set of observables whose values can be obtained via measurements
carried out within the domain  $\OO$. Such observables are called
local (localized in the domain $\OO$)~\cc{emch,hor}. Strictly
speaking, all observables should be considered as local. However,
global (quasi-local) observables are normally also considered in
the theory. They are constructed as limits of sequences of local
observables.

Using the localization feature, a physical system is
mathematically a set of observables localized in a domain.
However, other features should also be considered in practice. For
example, consider a solid body. Generally, a solid body is
characterized by a huge number of observables. Some of them
characterize the solid body as a whole. These observables can be
considered within the classical physics. Other observables
characterize individual molecules. These observables cannot be
described within the classical physics. At the same time, both
types of observables are localized in the domain  $\OO$, occupied
with the solid body. Thus, in this case the selecting feature of
the system is not only the localization of the observables, but
also their classical character.

This example demonstrates also that it may be possible to single
out from a quantum system characterized by a multitude of
compatible and incompatible observables a classical subsystem
characterized only with compatible observables. In this, isolation
of subsystem from the other part of the system is not assumed. For
example, elastic collisions are successfully described within
classical physics. In these processes, molecules described by
quantum physics are actively involved. Within the classical
description, participation of molecules can be taken into account
via boundary conditions, or via effective external field.

Let us discuss now the notion of the state of a physical system.
We start with discussion of classical systems. In this case, the
state is the attribute of the system that defines the outcomes of
all measurements of the observables unambiguously. From Newton's
times, the locality principle has been accepted in physics.  It
assumes, in particular, that the state of a localized physical
system is determined by the internal characteristics of the system
and the characteristics of the external fields acting on the
system that refer to the localization domain of the physical
system. By the classical paradigm, there exists a local reality
governing the state of the system.

Mathematically, a state is given by a point in a phase space. It
is assumed here that dynamics of the system is set. In the
approach of this paper, this way of setting the state is
inconvenient. First, it is hard to carry it over to the quantum
case. Second, it is rigidly associated with a particular way of
setting the dynamics. In particular, it assumes the introduction
of canonically conjugate variables. However, it is easily seen
that this way of setting a state is merely a particular way of
setting a real functional on the algebra of observables. This
functional is a character of the algebra. If we do not restrict
our consideration to a particular variant, a state of a classical
system can be defined as a character of the algebra of observables
of the system.

Let us now turn to the quantum case. The set  $\AAA_+$ of quantum
observables cannot be equipped with the structure of associative
commutative algebra. In view of this, it is not possible to
directly carry over the classical definition of the state to the
quantum case. However, it is possible to consider the set $\AAA_+$
as the totality of subsets $\QQQ$ $(\xi\in\Xi)$. And each subset
$\QQQ$ can be considered as a set of observables corresponding to
a classical subsystem of the quantum system. Certainly, such a
classical subsystem is not isolated form the rest of the system.
However, the possibility of selecting a subsystem is not strictly
stipulated by isolation from the rest of the system.

The state of each such subsystem is mathematically defined as
before via a functional  $\vx(\cdot)$ defined on the algebra
$\QQQ$, which is a character of this algebra. In this context we
introduce a new notion --- the notion of elementary state.

\

\df{} 39. The elementary state of a physical system is a set
$\vp=[\vx]$ ($\xi\in\Xi$) of functionals  $\vx$. Each of them is a
character of the corresponding algebra  $\QQQ$. The sets  $\QQQ$
are the maximal subsets of the set  $\AAA_+$ having the structure
of real associative commutative algebra.

\

The term "state" is justified also in the quantum case. Indeed, in
any particular measurement and even in a set of compatible
measurements, we can at best find values for a set of compatible
observables. All these observables belong to a single algebra
$\QQQ$. Therefore, their values are determined by the
corresponding functional $\vx$. All such functionals are defined
by elementary state $\vp$. Consequently, it fixes the results of
all possible measurements. In the standard quantum mechanics,
another mathematical object is associated with a state. In view of
this, we call $\vp=[\vx]$ an "elementary state." This term is not
used in the standard quantum mechanics. This stage of
considerations is concluded with the following postulate.

\

\post{} 4. The outcome of any particular experiment measuring the
observables of a physical system is determined by the elementary
state of the system.

\

The last postulate gives nothing new in the classical case. On the
contrary, it is a completely unconventional postulate in the
quantum case. Moreover, there are multiple proofs that nothing
like it can be true. This postulate is central to our approach. It
is possible to take that an elementary state realizes the concept
of "potential possibilities." This notion was introduced by
Fock~\cc{fok1}, but he did not give it a mathematical formulation.

Let us stress that we do not make extra assumptions on the
properties of the functionals  $\vx$. In particular, we do not
assume that
   \beq{6}
 \vx(\A)=\vp_{\xi'}(\A), \mbox{ if }\A \in \QQQ\cap\qqq.
\eeq

Certainly, equality \rr{6} may hold for some elementary states
$\vp$. We say that an elementary state  $\vp$ is stable on the
observable $\A$ if \rr{6} holds for any $\QQQ$ and $\qqq$
containing observable $\A$.

On the other hand, it seems to be very natural to require
fulfillment of equality  \rr{6}. In view of this, we have to
comment on the violation of this equality.

The experimental values of observables appear as responses of a
measuring device to the influences from the system under
measurement. Generally, responses of different devices to one and
the same influence may be different. From the standpoint of the
experimenter, such data is bad. The experimenter strives for
unification of the response (readings) of the devices. The
unification is achieved via device calibration.

Schematically, calibration goes as follows. A measuring device
serves as a reference one if it performs a reproducible
measurement of an observable $\A$. With this device, observabl
$\A$ is measured for a test physical system. It results in a value
$A$. By the definition of a reproducible measurement, repeated
measurement of the same observable with a device under calibration
should yield the same value. Only a device able to pass such a
test many times can be called a measuring device. Calibration is
to exclude dependence of the measurement outcome on the
uncontrollable influence of the device, in particular, on the
uncontrollable state of the device.

However, calibration is unable to exclude dependence on a
particular parameter whose value may be a property of the device.
This parameter is $\xi\in \Xi$. Let us clarify the relation
between parameter $\xi$ and the measuring device. Any device is
dedicated depending on its construction (tuning) either to the
measurement of a single observable $\A$, or to the simultaneous
measurement of a set of observables. This observable (set of
observables) belongs to an algebra $\QQQ$. We take that a device
is of type  $\xi$ if it is dedicated to measurements of an
observable (observables) belonging to the subset $\QQQ$, and,
secondly, if the outcome of the measurement of observable
$\A\in\QQQ$ (or set of compatible measurements) is
$A_{\xi}=\vx(\A)$ (set of corresponding results).

Calibration cannot ascertain if the outcome of measurement depends
on parameter~$\xi$ or not. Indeed, calibration starts with a
reproducible measurement. After this measurement, the state of the
system testing the device becomes stable on observab $\A$.
Therefore, the outcome of a subsequent measurement of this
observable will not in any case depend on parameter $\xi$. Any
other way of checking equality~\rr{6} implies that we should put
one and the same testing system under two measurements: once under
measurement by a device of type $\xi$, and again by a device of
type  $\xi'$ $(\xi\neq\xi')$. The devices are different, and, for
this reason, the measurements cannot be performed simultaneously.

Assume that the first measurement with the device of type $\xi$
has outcome  $A_{\xi}=\vx(\A)$. If this is an irreproducible
measurement, the elementary state $\vp$ of the system under
measurement will change uncontrollably. In this case, the outcome
of the second measurement (with device of type  $\xi'$) will not
be related in any way to the result of the first measurement. If
the first measurement is reproducible, elementary state  $\vp$
will be replaced with  $\vp'$ after the measurement. Since after a
reproducible measurement, the elementary state becomes stable on
the corresponding observable, the state  $\vp'$ should satisfy
relation $\vp'_{\xi'}(\A)=\vp_{\xi}(\A)$ regardless of the
fulfillment of equality~\rr{6}. We conclude that checking~\rr{6}
is beyond our reach in any case.

Certainly, the above considerations do not guarantee that the
outcomes of measurements do depend on  $\xi$. They only
demonstrate that such a dependence is possible. Therefore, any
conclusions based on assuming the validity of equality~\rr{6},
have no argumentativeness. We stress that type classification of
measuring devices by values of $\xi$ is a classification by the
character of interaction between the device and the system under
measurement. Therefore, not only the properties of the device, but
also the system under measurement (the set  $\AAA_+$ and algebras
$\QQQ$) determine it.

The dependence of measurement outcome on the device type can be
considered as a realization and concrete definition of Bohr's
views~\cc{bohr2} on the dependence of measurement outcome on the
general context of experiment. At the same time, the particular
variant of the dependence we suggest here does not contradict the
causality principle or the conception of local reality existence.
Local reality does not imply a definite value of any observable of
the physical system under consideration. It is rather a cause
effecting a particular reaction in a certain type of measuring
device.

Any measuring device is a classical system, and any external
influence on it can be considered as an effect of some external
field acting on the device. Here, the system under measurement
should be the source of the field. The device as a classical
system is insensitive to microscopic quantum details of the field,
only its classical characteristics matter. Therefore, in this
context, this field can be considered as an effective classical
field having characteristics governed by the elementary state of
the system under measurement. Namely, this effective field should
be considered as local reality dictating the outcome of any
particular measurement.

Only when the elementary state is stable on an observable is it
possible to speak of a definite value of this observable. In the
classical case, set $\Xi$ contains a single element. Thus, all
measuring devices belong to one and the same type.
Correspondingly, all elementary states are stable on any
observable, i.e., any observable has a definite value.

Measurement with a classical device cannot fix the elementary
state of a quantum system unambiguously. Indeed, as the devices
dedicated to measurements of incompatible observables are
incompatible, only observables belonging to a single algebra
$\QQQ$. are accessible to a measurement in a particular
experiment. As a result, only the values of the functional $\vx$.
will be established. The rest of the elementary state  $\vp$ will
remain undetermined. Repeated measurement employing a device of
another type will yield additional information, but will perturb
uncontrollably the elementary state established after the first
measurement. And the information obtained in the first measurement
will become useless.

In view of this, we take the following definition.

\

\df{} 40.  Elementary states $\vp$ are called $\vx$-equivalent if
their restrictions  $\vx$ on the algebra  $\QQQ$ are identical.

\

Thus, quantum measurement is only able to establish the
equivalence class of the physical state under measurement. The
procedure of preparation of a pure state in the standard quantum
mechanics can be easily recognized in the reproducible measurement
of observables belonging to algebra  $\QQQ$. In the following, we
call this the quantum state. The definition for it reads as
follows.

\

\df{} 41. Quantum state  $\Psi_{\vp\xi}$ is the class
$\{\vp\}_{\vp\xi}$ \quad $\vx$-equivalent elementary states that
are stable on algebra  $\QQQ$.

\

In fact, this definition is convenient only for systems without
identical particles. The problem is that the measuring device is
unable to determine which of the identical particles has been
detected. In view of this, a generalization of the definition is
convenient. We will say that elementary state  $\vp$ is weakly
$\vx$-equivalent to elementary state  $\vp'$ if restriction  $\vx$
of the elementary state  $\vp$ onto algebra  $\QQQ$ coincides with
restriction  $\vp'_{\xi'}$ of the elementary state $\vp'$ on the
algebra  $\qqq$ that results from   $\QQQ$ after replacement of
the observables of one of the identical particles with the
corresponding observables of another identical particle. For
systems with identical particles, the equivalence involved in the
definition of the quantum state should be replaced with a weak
equivalence. We assume hereafter that such replacement is
performed if necessary.

In this way, our quantum state characterizes not a single state
but an ensemble of such objects. We share this view with
Blokhintsev~\cc{blo1,blo2,blo3,blo4}.

\section{PROBABILITY THEORY AND QUANTUM ENSEMBLE}

Most predictions of quantum theory are probabilistic. Therefore,
quantum theory should be based on probability theory. Presently,
Kolmogorov probability theory is the most developed
mathematically~\cc{kol}. It is commonly believed that quantum
systems require dedicated quantum probability theory. Here, we
advocate the view that classical Kolmogorov probability theory is
quite adequate to the quantum case if one takes into account the
peculiarities of quantum measurements~\cc{slav4}.

The basic notion of Kolmogorov probability theory (see,
e.g.,~\cc{kol,nev}) is the so-called probability space  $(\Om,\F,
P)$.

The first component  $\Om$ is a set (space) of elementary events.
The physical meaning of elementary events is not specified, but it
is assumed that the events are mutually exclusive, and one and
only one event is realized in each trial. In our case an
elementary event appears as an elementary state  $\vp$.

\

\rem. Evidently, it is not possible to use a quantum state as an
elementary event, because two nonorthogonal events are not
mutually exclusive. Therefore, it is indeed not possible to use
Kolmogorov probability theory within the standard quantum
mechanics. The classical formal logic does not fit in the standard
quantum mechanical framework for similar reasons.

\

Apart form the elementary event, the notion of a random event, or,
for brevity, event (without any adjective) is introduced. Any
event  $F$ is identified with a subset of the set $\Om$. Event $F$
is assumed to be realized if any of the elementary events
belonging to this subset ($\vp\in F$) has realized itself. It is
assumed that we can establish whether an event is realized or not
for any trial. There is no need to assume the same about
elementary events.

Sets of subsets of the set  $\Om$ (including the set  $\Om$ itself
and the empty set  $\emptyset$) are equipped with the structure of
Boolean algebra. Correspondingly, the second component of
probability space is a Boolean $\sss$-algebra  $\F$. In this way,
probability space is endowed with the structure of measurable
space.

Finally, the third component of probability space is a probability
measure  $P$. It is a mapping of the set  $\F$ onto the set of
real numbers (each  $F\in\F$ is mapped to a number $P(F)$), This
mapping satisfies the following conditions: (a)  $0\leq P(F) \leq
1$ for any  $F\in\F$, $P(\Om)=1$; (b) $P(\sum_j F_j)=\sum_j
P(F_j)$ for any countable family of disjoint subsets  $F_j\in \F$.

We stress that probability measure is defined only for the events
belonging to algebra  $\F$. Generally, the probability may not be
defined for elementary events.

Let us clarify the last statement with an example. Let the space
of elementary events be the set of rational numbers lying between
zero and unity. A trial is guessing of a number given by an
interlocutor. Evidently, the probability of guessing of a number
cannot have any value other than zero. However, it also cannot be
zero. Indeed, any given number is with unit probability between
zero and unit. The set of rational numbers is countable. Thus, by
the properties of probability measure, a unit should be a
countable sum of zeros. Contradiction is avoided if we take as
$\F$ the set of all intervals (and their unions), and set the
probability of each interval to its length.

We see that measurability is a crucial property of probability
space. We will see below that it is even more important in the
quantum case. Also, measurability is needed not only for
mathematical consistency; it carries an important physical
meaning.

Let us discuss now how the basic principles of probability theory
are applied to quantum measurements. Most quantum measurements are
related to obtaining probability distributions of some
observables. With particular measuring devices we can obtain such
distribution for a collection of compatible observables. In terms
of probability theory, choosing a certain measuring instrument we
chose a specific  $\sss$-algebra $\F$.

For clarity, we continue the discussion with an example. Let the
system under study be a particle that moves in a fixed plane. Let
us first look for a probability distribution of coordinate  $X$ of
the particle. Aiming at this, we split the plane into strips
perpendicular to axis $X$. The width of the strips should be
adjusted to the sensitivity of the measuring device. The strips
will be elements $F^X_i$ of the algebra  $\F_X$. With the
measuring device we can determine the probability for the particle
to be detected in a particular strip. Similar experiment can be
performed to determine the probability distribution over the axis
$Y$. In this case, the strips will be denoted as $F^Y_j$, and the
$\sss$-algebra as  $\F_Y$.

We can perform a more detailed study and obtain the probability
distribution in both coordinates simultaneously. To this end, we
are to split the plane into rectangles resulting from
intersections of various strips: $F^{XY}_{ij}=F^X_i\cap F^Y_j$.
The rectangles  $F^{XY}_{ij}$ will be the elements of
$\sss$-algebra  $\F_{XY}$. Algebra $\F_{XY}$ is referred to as
being generated by the algebras $\F_X$ and $\F_Y$. Until now,
there was no difference between classical and quantum
considerations.

Suppose now that we are looking for probability distributions not
only for coordinates but also for momenta. If we are interested in
distributions over coordinates and separately over momenta, the
experiment can be organized as before. The only difference is that
the strips should now be in the plane of momenta.

The situation is drastically different if we try to determine the
combined probability distribution over the  $X$th coordinate and
$K_x$th projection of the momentum. Formally, in mathematics (see,
e.g.,~\cc{nev}), we can construct a  $\sss$-algebra $\F_{XK_x}$
generated by the algebras $\F_{X}$ è $\F_{K_x}$. The elements of
this algebra are rectangles (and various unions of the rectangles)
in the two-dimensional plane  $(XK_x)$ of the four-dimensional
phase space. In the classical case, we can perform an experiment
to determine the probability of detecting a particle in such a
rectangle. However, in the quantum case, such an experiment is not
possible because the devices measuring the  $X$th coordinate and
$K_x$th projection of the momentum are not compatible. This
implies that it is not possible to ascribe a probability measure
to such a rectangle. In other words, there is no notion of
probability for an event of detecting a particle in such a
rectangle.

From this example, we draw the following general conclusion. Not
any mathematically possible (and classically allowable)
$\sss$-algebra is allowable as the  $\sss$-algebra of probability
space in quantum space.

We conclude that experimentally an element of the measurable space
$(\Om,\F)$ corresponds to a pair of the object under study and a
particular type of measuring device. This type of device allows
one to register an event corresponding to a set of compatible
measurable quantities, i.e., corresponding to a fixed algebra
$\QQQ$. In view of this, $\sss$-algebra  $\F$ can also be labeled
with parameter  $\xi$: $\F=\F_{\xi}$.

The specifics of quantum experiments require particular care in
defining one of the basic notions of probability theory, the real
random quantity. Customarily, the real random quantity is defined
as a mapping of space  $\Om$ of elementary events onto the
extended real axis  ${\cal R}=[-\infty,+\infty]$. However, this
definition does not take into account the specifics of the quantum
experiment, where the outcome may depend on the type of measuring
device. In view of this, we take the following extended
definition.

\

\df{} 42.   A real random quantity is a mapping of a measurable
space  $(\Om,\F_{\xi})$ of elementary events onto the extended
real axis.

\

For observable  $\A$ it can be expressed as follows:
 $$
\vp\stackrel{\A}{\longrightarrow}A_{\xi}(\vp)\equiv\vx(\A)
\in{\cal R}.$$

We call the quantum ensemble a set of identical physical systems
(this means that they share the set  $\AAA_+$ of observables, and
the set  $\{\QQQ\}$ of commutative algebra $\QQQ$ ($\xi\in\Xi$))
that are in some quantum state. A quantum ensemble mixture that
involves each of these ensembles with multiplicity $C_i$
($C_i\geq0, \quad \sum C_i<\infty$) is called a mixed ensemble.
Experiment testifies in favor of the following postulate.

\

\post{} 5. A quantum (generally, mixed) ensemble can be equipped
with the structure of probability space. A reproducible
measurement transforms a quantum ensemble into a new quantum
ensemble that may have a new probability distribution of
observables.

\

Consider an ensemble of physical systems occupying the quantum
state  $\Psi_{\vp\eta}$ $(\eta \in \Xi)$. In this case, we
consider the equivalence class  $\{\vp\}_{\vp\eta}$ as the space
$\Om(\vp_{\eta})$ of elementary even $\vp$. Let a value of
observable  $\A \in \QQQ$ be measured in an experiment employing a
device of type $\xi$. We denote as  $(\Om(\vp_{\eta}),\F_{\xi})$
the corresponding measurable space. Let  $P_{\xi}$ be the
probability measure on this space; i.e., $P_{\xi}(F)$ is the
probability of event  $F\in\F_{\xi}$.

We take that event  $\aA$ is realized in an experiment if the
detected value of observable  $\A$ does not exceed  $A$. The
probability of this event is  $P_{\xi}(\aA)=P(\vp:
\vp_{\xi}(\A)\le A)$. Knowing the probabilities  $P_{\xi}(F)$
suffices to obtain via summations and integrations the probability
$P_{\xi}(\aA)$; distribution  $P_{\xi}(\aA)$ is marginal with
respect to the probabilities  $P_{\xi}(F)$ (see, e.g.,~\cc{pro}).

Observable $\A$ may belong, apart from the algebra  $\QQQ$, to
another maximal algebra  $\qqq$. If so, we can use a device of
type $\xi'$  to measure the probability of the event  $\aA$. In
this case, another value $P_{\xi'}(\aA)$ could be obtained for the
probability. However, experiment shows that the probabilities do
not depend on the measuring device in use. Thus, we should adopt
one more postulate.

\

\post{} 6.  If observable  $\A\in\QQQ\cap\qqq$ the probability of
detecting event  $\aA$ for a system in quantum state
$\Psi_{\vp\eta}$ does not depend on the type of measuring device,
i.e., $P(\vp: \vpx)\le A)= P(\vp:\vp_{\xi'}(\A)\le A)$.

\

Thus, despite the fact that elementary state  $\vp$ is a set of
functional  $\vx$, we can use the notation $P(\vp:\vp(\A)\le A)$
for the probability of event  $\aA$.

Let us also introduce the following notation:
 $$P_{\A}(d\vp)=P(\vp:\vp(\A)\leq A+dA)-P(\vp:\vp(\A)\leq A)$$
and consider the ensemble of quantum systems occupying the quantum
state  $\Psi_{\vp\eta}$. By probability theory (see,
e.g.,~\cc{nev}) the expectation value of observable $\A$ in this
state is given by the expression
 \beq{10}
\langle\A\rangle=\int_{\vp\in\Psi_{\vp\eta}}\,P_{\A}(d\vp)\,A(\vp)
\equiv\int_{\vp\in\Psi_{\vp\eta}}\,P_{\A}(d\vp)\,\vp(\A).
  \eeq

On the other hand, the Khinchin theorem holds (see the law of
large numbers, for example, in~\cc{nev}):

\

{\sc Theorem.}  Let $A_i=\vp_i(\A) \qquad (1\leq i \leq n, \quad
\vp_i\in\Psi_{\vp\eta} )$ be a sequence of mutually independent
randomly selected quantities with one and the same probability
distribution having a finite expectation value $\langle\A\rangle$.
Then the quantity   $(A_1+\dots+A_n)/n$ converges in probability
to $\langle\A\rangle$ at $n\to\infty$. Thus,
 \beq{11}
 \Psi_{\vp\eta}(\A)\equiv\lim_{n\to\infty}\mbox{P}\Big[n^{-1}
\Big(\vp_1(\A)+\dots+\vp_n(\A)\Big)\Big] = \langle\A\rangle.
  \eeq

Equation~\rr{11} defines a functional (the quantum average) on the
set  $\AAA_+$. We denote this functional as
$\Psi{\vp\eta}(\cdot)$, and call it also the quantum state.
Equation~\rr{11} and the properties of functionals  $\vp_i(\cdot)$
imply immediately that $\Psi{\vp\eta}(\cdot)$ is linear on each
subset $\QQQ$ of compatible observables. In other words,
restriction of functional $\Psi{\vp\eta}(\cdot)$ onto each subset
$\QQQ$ is a linear functional. The linearity property of
functional  $\Psi{\vp\eta}(\cdot)$ can be extended on the whole
set  $\AAA_+$. Before this, the set $\AAA_+$ should be equipped
with the structure of real linear space.

Since any element  $\A$ of the set  $\AAA_+$ belongs to some
linear space  $\QQQ$, its multiplication by real numbers is
defined. The addition of elements $\A$ and $\B$ is more involved,
because these elements may belong to different linear spaces
$\QQQ$ and $\QQ_{\xi'}$. However, the totality of quantum
experiments points to the following fact. For any $\A$ and $\B$
belonging to  $\AAA_+$ there exists such  $\D\in\AAA_+$ that for
any quantum state  $\Psi{\vp\eta}(\cdot)$ we have

 $$\Psi{\vp\eta}(\A)+\Psi{\vp\eta}(\B)=\Psi{\vp\eta}(\D).$$

 This element  $\D$ can be taken by definition to be the
sum of elements $\A$ and $\B$, i.e., $\D=\A+\B$. In view of these
considerations, we accept the following postulate.

 \

\post{} 7. The set  $\AAA_+$ can be equipped with the structure of
a real vector space, and functionals
 $\Psi{\vp\eta}(\cdot)$ are linear on that space.
 \

 This means that
 $$\Psi{\vp\eta}(\A)+\Psi{\vp\eta}(\B)=\Psi{\vp\eta}(\A+\B)$$

This relation also holds when  $\A$ and $\B$ belong to different
subsets   $\QQQ$ and $\QQ_{\xi'}$.

The set  $\AAA_+$ can be equipped with the structure of real
algebra. The product is defined as follows:
 \beq{12}
 \A\circ\B=1/2\lt((\A+\B)^2-\A^2-\B^2\rt).
  \eeq

This product is obviously commutative, but may be nonassociative
(see (D.3)), i.e., associator  $\{\A,\B,\D\}=(\A\circ
\B)\circ\D-\A\circ(\B\circ\D)$ is not necessarily zero. It can be
demonstrated (see~\cc{emch}) that vanishing of the associator
$\{\A,\B,\al\I\}$ for any $\A$ and $\B$ and any real number $\al$
is necessary and sufficient for distributivity (see D.2.b,c) of
the product  $\A\circ\B$. Under this condition, the real algebra
with the product~\rr{12} is called real Jordan
algebra~\cc{jord,emch}.

In principle, it is possible to base quantum theory on this
algebra. Success has been very limited in this direction
(see~\cc{emch}). Much more successful has been the approach based
on a complex associative algebra, which, in a certain sense,
contains the Jordan algebra as it real part.

All Jordan algebras are classified into special and exceptional.
The special algebras are defined as follows. Consider real or
complex algebra \AAA{} with a "conventional" product  $\U\V \qquad
(\U\in\AAA,\V\in\AAA,\U\V\in\AAA)$. The algebra is associative
with respect to this product, but not necessarily commutative. One
can introduce the "symmetrized" product for the set  \AAA{}
 \beq{13}
 \U\circ\V=1/2(\U\V+\V\U).
  \eeq

With respect to this product, the set  \AAA{} is a Jordan algebra.
Any Jordan algebra isomorphic to such an algebra (or to its
subalgebra) is called a special algebra. Otherwise, a Jordan
algebra is called exceptional. Not any Jordan algebra is special.
Thus, to be special, a Jordan algebra should imply some identities
between its elements. In principle, such identities could be
checked in experiments. However, the list of these identities is
presently not known. On the other hand, in all quantum mechanical
models considered until now the set of observables can be equipped
with the structure of a special Jordan algebra. We stay within
this tradition and accept the following hypothesis.

   \

{\sc Hypothesis. } There exists an involutive, associative,
probably noncommutative algebra  \AAA{} satisfying the following
conditions:\\
  (a) for any element  $\U\in\AAA$ there exists a Hermitian element
  $\A$ such that  $\U^*\U=\A^2$;\\
  (b) $\U^*\U=0$ implies that  $\U=0$;\\
  (c)  the set of Hermitian elements of the algebra  \AAA{} coincides
  with the set  $\AAA_+$ of the observables.

  \

Hereafter, elements of algebra  \AAA{} are called dynamical
quantities.

Evidently, set  $\AAA_+$ can be equipped with the structure of
Jordan algebra defining the product of its elements with
equation~\rr{13}. This hypothesis means that the Jordan algebra of
observables is special and real. Correspondingly, dynamical
quantities can be added and multiplied using the usual rules of
addition and multiplication (apart from commutation). This seems
to be so evident that it is almost never discussed. Despite this,
we call the corresponding statements a hypothesis, not a
postulate, because we cannot point out an experimental means of
checking the necessity of these statements.

We stress that in the standard quantum mechanics the statements of
the hypothesis are accepted in a much stronger form. Apart from
our assumptions, it is assumed there that the observables are
self-adjoint operators in a Hilbert space. These extra assumptions
are hardly self-evident.

Hereafter, we consider a physical system to be given if the
algebra  \AAA{} of its dynamic quantities is given. By the first
postulate, the algebras  $\QQQ$ of compatible observables are the
maximal real commutative subalgebras of algebra \AAA{} belonging
to  $\AAA_+$. This implies in turn that the compatible observables
are mutually commutative elements of algebra \AAA{} and the
incompatible observables do not commute with each other.

We mentioned above that in the quantum case, the set  $\Xi$ of
subalgebras  $\QQQ \quad (\xi\in\Xi)$ has cardinality of
continuum. Indeed, even if algebra  \AAA{} is an algebra with only
two noncommuting generators $\A_1$ and $\A_2$, the commutative
algebra  $\QQ_{\al}$ with the generator  $\A_{\al}=\A_1
\cos\al+\A_2\sin\al$ will be an algebra of type  $\QQQ$ at any
real  $\al$.

\section {$C^*$-ALGEBRA AND HILBERT SPACE }

Any element  $\U$ of algebra  \AAA{} can be uniquely represented
in the form  $\U=\A+i\B$, where $\A,\B\in\AAA_+$. Therefore,
functional  $\Psi_{\vp\eta}(\,\cdot\,)$ can be uniquely extended
to a linear functional on algebra \AAA:
$\Psi_{\vp\eta}(\U)=\Psi_{\vp\eta}(\A)+i\Psi_{\vp\eta}(\B)$.

Let us define a semi-norm of element $\U$ with the
  \beq{14}
 \|\U\|^2=\sup_{\xi}\,\sup_{\vx}\vx(\U^*\U)=r(\U^*\U),
 \eeq
where $r(\U^*\U)$ is the spectral radius of the element  $\U^*\U$
on algebra  \AAA.

This is an acceptable definition. First,  $\|\U\|^2\ge 0$ due to
the property (P.7.c). Next, due to definition of the probability
measure, we have for any  $\eta\in \Xi$ that
 \beq{15}
\Psi_{\vp\eta}(\U^*\U)=\int_{\vp \in\{\vp\}_{\vp\eta}}\,P_{\U^*\U}
(d\vp)\,\vp(\U^*\U)\le\sup_{\xi}\,\sup_{\vx}\vx(\U^*\U)=r(\U^*\U).
 \eeq
Next, for such  $\eta \in \Xi$ that $\U^*\U\in\QQ_{\eta}$ we have
$\Psi_{\vp\eta}(\U^*\U)=\vp_{\eta}(\U^*\U)$. Thus, for such $\eta$
 \beq{16}
\sup_{\vp_{\eta}}\Psi_{\vp\eta}(\U^*\U)=\sup_{\vp_{\eta}}\vp_{\eta}
(\U^*\U)=r_{\eta}(\U^*\U),
  \eeq
where $r_{\eta}(\U^*\U)$ is the spectral radius in  $\QQ_{\eta}$
since the subalgebra  $\QQ_{\eta}$ is maximal (see (P.2),
$r_{\eta}(\U^*\U)=r(\U^*\U)$. From this, using Eqs. \rr{14},
\rr{15}, and \rr{16}, we obtain
 \beq{17}
 \|\U\|^2=\sup_{\xi}\,\sup_{\vx}\vx(\U^*\U)=\sup_{\xi}\,\sup_{\vx}\Psi_{\vp\xi}(\U^*\U).
 \eeq

Since $\Psi_{\vp\xi}(\,\cdot\,)$ is a linear positive functional,
the Cauchy-Schwarz inequality holds (see (P.5.b)):
 \beq{18}
|\Psi_{\vp\xi}(\U^*\V)\Psi_{\vp\xi}(\V^*\U)|\le
\Psi_{\vp\xi}(\U^*\U)\Psi_{\vp\xi}(\V^*\V).
 \eeq

This implies that  $\|\U\|$ satisfies the semi-norm axioms for
element $\U$ (see, e.g., \cc{emch}): $$\|\U+\V\|\le\|\U\|+\|\V\|,
\quad \|\lll\U\|=|\lll|\|\U\|, \quad \|\U^*\|=\|\U\|, \quad
\|\U\V\|\leq\|\U\|\|\V\|.$$

Let us consider now the set  $\JJ$ of elements  $\U$ of algebra
\AAA{} for which  $\|\U\|^2=0$. Inequality~\rr{18} implies that
$\JJ$ is a two-sided ideal of  \AAA. Thus, we can form a factor
algebra $\AAA'=\AAA/\JJ$. For algebra  $\AAA'$,  $\|\U\|^2=0$
implies that  $\U=0$. Thus, equality~\rr{14} defines a norm on
algebra  $\AAA'$. On the other hand, it can be checked that
algebra $\AAA'$ contains the same physical information as \AAA{}
does.

For this check, we consider two observables  $\A$ and $\B$ that
either simultaneously belong or simultaneously do not belong to
each of the subalgebras  $\QQQ$. Let $\A$ and $\B$ satisfy the
extra condition $\|\A-\B\|=0$. Then equation~\rr{14} implies that
  \beq{19}
  \vpx=\vx(\B)
  \eeq
for all  $\QQQ$ containing these observables. Equality~\rr{19}
means that no experiment can tell the difference between these
observables. Thus, from the phenomenological standpoint, these
observables should be considered as identical. Mathematically,
these observables are equivalent modulo the ideal  $\JJ$. Under
the transition from algebra  $\AAA$ to algebra  $\AAA'$, all the
equivalent observables are mathematically equated. Making a
shortcut to algebra $\AAA'$, we accept the following postulate.

\

\post{} 8.  If $\sup_{\xi}\,\sup_{\vx}|\vx(\A-\B)|=0,$  $\A=\B$.

\

This postulate is of a technical character. At the same time, it
does not pose any restrictions from the phenomenological
standpoint. Its only role is to simplify the mathematical
description of physical systems. Hereafter, we assume that
Postulate 8 is fulfilled, and, therefore, Eq.~\rr{17} defines the
norm of element  $\U$.

The multiplicative properties of functional $\vx$ imply that
$\vx([\U^*\U]^2)=[\vx(\U^*\U)]^2$. Therefore,
$\|\U^*\U\|=\|\U\|^2$. Thus, completion of algebra  \AAA{} by the
norm  $\|\cdot\|$ makes \AAA{} a $C^*$-algebra~\cc{dix}. We
conclude that the algebra of quantum dynamical quantities can be
equipped with the structure of $C^*$-algebra. In the standard
algebraic approach to quantum theory, this statement plays the
role of a basic axiom. Mathematically, it is very convenient.
However, phenomenologically, the necessity of this axiom was
unclear.

In most of our previous constructions, elementary state
$\vp=[\vx]$ was a basic ingredient. The elementary state possesses
many of the properties usually ascribed to the so-called hidden
parameters~\cc{bohm1}. Since the times of von Neumann~\cc{von},
there is a strong opinion in standard quantum mechanics that
hidden parameters cannot exist in quantum mechanics. In view of
this, it is necessary to verify that elementary states can be
introduced without contradictions.

Setting a physical system assumes setting an algebra  \AAA{} of
dynamical quantities. As we have argued above, this algebra should
have the structure of  $C^*$-algebra. Setting algebra  \AAA, we
set at the same time the set of its maximal commutative
associative subalgebras  $\QQQ$. Each of these subalgebras is a
Banach algebra.

Clearly, constructing any elementary state $\vp=[\vx]$ is
equivalent to constructing all its components  $\vx$, and each
$\vx$ is a character of the subalgebra $\QQQ$. Each functional
$\vx$ can be constructed as follows. Select in an arbitrary way
the system of independent generating elements $G(\QQQ)$ in the
subalgebra  $\QQQ$. Using (P.8), relate each element of the set
$G(\QQQ)$ to a point of its spectrum. In this way, we define the
functional $\vx$ on the set  $G(\QQQ)$. By linearity and
multiplicativity, functional  $\vx$ can be extended unambiguously
to the whole subalgebra $\QQQ$. Sorting out all the points of the
spectrum for each element of the set  $G(\QQQ)$, we construct all
possible functionals  $\vx$. For another  $\xi$, functionals $\vx$
are constructed in the same way. This is a consistent procedure if
the functionals are constructed independently for different $\xi$.
If we impose the condition~\rr{6}, the procedure may turn out, and
sometimes does turn out self-contradictory.

However, it is always possible to construct an elementary state
$\vp$ that is stable on all the observables belonging to any
single subalgebra  $\QQQ$. This is achieved if we start
constructing the functional  $\vp$ from this very subalgebra using
the above procedure. On another subalgebra $\qqq$, the functional
$\vp_{\xi'}$ is constructed as follows. Let  $\QQQ\cap
\qqq=\QQ_{\xi\xi'}$, and let $G(\QQ_{\xi\xi'})$ be the independent
generating elements of subalgebra $\QQ_{\xi\xi'}$. Let $\tilde
G(\QQ_{\xi\xi'})$ be a supplement of these generating elements to
the set of generating elements of algebra $\qqq$. If $\A\in
\QQ_{\xi\xi'}$, set $\vp_{\xi'}(\A)=\vx(\A)$. If $\A\in \tilde
G(\QQ_{\xi\xi'})$, construct $\vp_{\xi'}(\A)$ as a mapping of the
element $\A$ to one of the points of its spectrum. Functional
$\vp_{\xi'}$ is extended to another element of subalgebra $\qqq$
by linearity and multiplicativity.

We conclude that there is no problem of existence for elementary
states.

The proof due to von Neumann~\cc{von} of the inexistence of hidden
parameters is invalid for the elementary states  $\vp$ for the
following reason. It is assumed in the proof that the state is
described by a linear functional on the set of observables.
Elementary state  $\vp$ can be considered as a functional on the
set of observables. However, it is linear only on the subsets
$\QQQ$. Besides, it is a multivalued functional.

In the same proof, von Neumann demonstrated that linearity of the
functional describing the state of the system is incompatible with
the assumption on the presence of microcausality. He concluded
that microcausality is absent, and macrocausality appears due to
averaging over a large number of noncausal events. Similarly,
Blokhintsev maintained the view~\cc{blo1,blo3,blo4} that causality
is not a property of individual quantum objects, but characterizes
only ensembles of such objects. It allows one to consider quantum
processes governed by the Schrodinger equation as causal. In our
approach we resolve the "linearity-causality" dilemma in the
opposite way. Microscopically, causality is present, but there is
no linearity of the state describing an individual quantum system.
Linearity of a quantum state appears due to averaging over a
quantum ensemble.

We point out that the appearance of linearity after averaging is a
not an infrequent phenomenon in probability theory. Therefore, the
principles of linearity and superposition considered as basic
physical principles in the standard quantum mechanics are deprived
of this status in reality. These properties are only mathematical
artifacts appearing due to an averaging procedure. In contrast,
causality is a physical principle, which is widely used in
physics, ignoring the "official ban." Namely, an elementary state
is able to claim the role of a mathematical image of the reality
physically supporting causality.

The superposition property mentioned above originates from the
following remarkable property of  $C^*$-algebra. Any $C^*$-algebra
is isometrically isomorphic to the subalgebra of linear bounded
operators in a Hilbert space~\cc{dix}. It will allow us later on
to use the familiar formalism of Hilbert space, where the
superposition property appears naturally.

\

\rem. It is usually assumed in the standard quantum mechanics that
all self-adjoint bounded operators in Hilbert space are
observables. This assumption is violated in the models with
superselection rules~\cc{wigh}. The algebraic approach (also, its
version in this paper) does not need this assumption.

\

In the algebraic approach, a state is defined as a positive linear
functional  $\Psi$ on the set of observables satisfying the
normalization condition  $\Psi(\I)=1$. In the standard quantum
mechanics, a state is given either with a vector of Hilbert space,
or, more generally, with a density matrix. However, not any
physically interesting state can be given with a density matrix
(see~\cc{emch}). Because of this, the algebraic definition is more
general. Frequently, a state defined in this way is called
algebraic. Since $\vp(\I)=1$ (see (P.7.b)), the functional
$\Psi_{\vp\eta}(\A)$ defined by formula~\rr{10} satisfies the
normalization condition. Thus, the quantum state defined in this
paper is an algebraic state. Since any linear positive functional
defined on the set of observables can be unambiguously extended on
the algebra of dynamical quantities, hereafter we call the
algebraic state a linear positive normalized functional defined on
algebra  \AAA.

\

\df{} 43. Algebraic state  $\Psi$ is called a pure state if the
equality
 \beq{21}
\Psi=\lll\Psi_1+(1-\lll)\Psi_2 \qquad 0<\lll<1,
  \eeq
 where $\Psi_1$ and $\Psi_2$ are two states, implies that
  $\Psi_1=\Psi_2$.

  \

It is easily checked that the quantum state $\Psi_{\vp\xi}$
introduced in definition (D.41) is a pure algebraic state. Indeed,
assume that functional  $\Psi_{\vp\xi}$ can be represented in the
form~\rr{21}. Restrict the equality~\rr{21} to the subalgebra
$\QQQ$. On this subalgebra, i.e., for any  $\A\in\QQQ$, the
equality $\vx(\A)=\Psi_{\vp\xi}(\A)$is valid. However, functional
$\vx(\cdot)$ is a character of subalgebra  $\QQQ$. Any character
of a commutative algebra is a pure state (see~\cc{brat}).
Therefore, equation
$\Psi_{\vp\xi}(\A)=\lll\Psi_1(\A)+(1-\lll)\Psi_2(\A)$ implies that
$\Psi_1(\A)=\Psi_2(\A)=\Psi_{\vp\xi}(\A)=\vx(\A)$ for any
$\A\in\QQQ$. In particular,
 $$\Psi_1([\A-\vx(\A)]^2)=\vx([\A-\vx(\A)]^2)=0.$$

 This implies that
 $$\int_{\vp\in\Psi_1} P_{\A}(d\vp)\,\vp([\A-\vx(\A)]^2)
\equiv\Psi_1([\A-\vx(\A)]^2)=0.$$ for any $\A\in\QQQ$. Thus, if
$\vp\in\Psi_1$, almost surely, $\vp(\A)=\vx(\A)$ for $\A\in\QQQ$.
This means that almost surely elementary states $\vp\in\Psi_1$
form the equivalence class $\{\vp\}_{\vp\xi}$. From this, we
obtain that
 $$\Psi_1(\A)=\int_{\vp\in\Psi_1}P_{\A}(d\vp)\,\vp(\A)=
\int_{\vp\in\Psi_{\vp\xi}}P_{\A}(d\vp)\,\vp(\A)=\Psi_{\vp\xi}(\A)$$
for any  $\A$. For $\Psi_2(\A)$, we have an analogous derivation,
i.e.,  $\Psi_1(\A)=\Psi_2(\A)$.

There is a procedure that realizes the relation between
$C^*$algebra and Hilbert space. It is called the
Gelfand-Naimark-Segal (GNS) canonical construction (see,
e.g.,~\cc{naj,emch}). Below, we give a brief exposition.

Suppose that some  $C^*$-algebra \AAA{} and a positive functional
$\po$ on this algebra are given. Two elements $\U,\,\U'\in\AAA$
are taken to be equivalent if the equality
$\po\left(\W^*(\U-\U')\right)=0$ holds for any  $\W\in\AAA$.
Denote by $\Phi(\U)$ the equivalence class containing element
$\U$, and consider the set of all equivalence classes $\AAA(\po)$
in \AAA. The set  $\AAA(\po)$ is a linear space if we take
$a\Phi(\U)+b\Phi(\V)=\Phi(a\U+b\V)$. The scalar product in
$\AAA(\po)$ is defined as follows:
  \beq{20}
\left(\Phi(\U),\Phi(\V)\right)=\po(\U^*\V).
  \eeq

This scalar product generates in  $\AAA(\po)$ the norm
 $\|\Phi(\U)\|=[\po(\U^*\U)]^{1/2}$. Completion by this norm
make $\AAA(\po)$ a Hilbert space. Each element  $\V$ of algebra
\AAA{} is uniquely represented in this space by a linear operator
 $\Pi(\V)$ acting by the rule
  \beq{22}
\Pi(\V)\Phi(\U)=\Phi(\V\U).
  \eeq

In this way, the GNS construction allows one to construct a
representation of any  $C^*$-algebra. Let us recall which
representations are possible.

Representations can be exact or inexact. For an exact
representation, different elements of the algebra are represented
by different operators in the Hilbert space.

\

\df{} 44. Representation $\V\to\Pi(\V)$ is called exact if
$\Pi(\V)=0$ implies $\V=0$.

\

A representation can be zero.

\

\df{} 45. Representation $\V\to\Pi(\V)$ is called zero if
$\Pi(\V)=0$ at any~$\V$.

\

\df{} 46. Representation $\V\to\Pi(\V)$ is a direct orthogonal
sum, $\Pi(\V)=\Pi_1(\V)\bigoplus\Pi_2(\V)$, of two (or a larger
number) representations if the operators of $\Pi(\V)$ act in the
Hilbert space $\HHH=\HHH_1\bigoplus\HHH_2$ by the rule
$\Pi(\V)\Phi=\Pi_1(\V)\Phi_1+\Pi_2(\V)\Phi_2$. Here,
$\Phi=\Phi_1+\Phi_2, \quad \Phi_1\in\HHH_1,\Phi_2\in\HHH_2$, while
$\Pi_1(\V)$ and $\Pi_2(\V)$ are the operators of representations
in the spaces  $\HHH_1$ and $\HHH_2$, respectively.

\

\df{} 47. Representation $\V\to\Pi(\V)$is called degenerate if it
is representable as a direct orthogonal sum of representations
among which at least one is a zero representation.

\

\df{} 48. Representation $\V\to\Pi(\V)$ is called irreducible if
it is not possible to represent it as a direct orthogonal sum of
two other representations.

\

\df{} 49. Representation $\V\to\Pi(\V)$ acting in a Hilbert space
$\HHH$ is called cyclic if there exists a vector $\Phi$ (called
the cyclic vector) such that the set of vectors $\Pi(\V)\Phi$ is
everywhere dense in $\HHH$.

\

The latter is equivalent to the requirement that  $\Pi(\V)\Phi$
contains a basis of  $\HHH$.

Evidently, GNS construction yields a cyclic nondegenerate
representation. It can be demonstrated that this representation is
irreducible if and only if  $\po$ is a pure state. Generally, this
representation is not an exact one. However, there exists the
so-called universal representation $\V\to\Pi_u(\V)$. This
representation is a direct sum of representations,
$\Pi_u(\V)=\bigoplus_i\Pi_i(\V)$. Each of the representations
$\V\to\Pi_i(\V)$ is yielded by GNS construction with a state
$\Psi_i$. Summation runs over all the algebraic states $\Psi_i$.

Any nondegenerate representation of a  $C^*$-algebra is isomorphic
to a subrepresentation of the universal representation. The
universal representation is exact. This means that the algebra of
elements $\V$ is isomorphic to the algebra of operators
$\Pi_u(\V)$. In other words,  $C^*$-algebra \AAA{} is isomorphic
to the subalgebra of bounded linear operators in the Hilbert space
$\HHH_u$. Any algebraic relation between elements of \AAA{} can be
established via establishing corresponding relations between the
operators realizing any exact representation of the algebra. The
existence of the universal representation guarantees that there
exists at least one such representation.

We mentioned above that the quantum states introduced in this
paper are pure algebraic states. Now we demonstrate how to
construct the functionals with the required properties. Let us
first consider the case when the commutative algebra  $\QQQ$,
defining the quantum state contains a one-dimensional projector
$\p_0$. The most accessible definition of the one-dimensional
projector is as such element of the algebra that is represented in
any exact representation with a projection operator onto
one-dimensional subspace of the Hilbert space.

\

\rem. In the standard quantum mechanics, it is usually assumed
that any bounded self-adjoint operator corresponds to an
observable. In this case, any maximal commutative subalgebra
contains one-dimensional projectors. Conversely, each
one-dimensional projector belongs to some commutative subalgebra.
In this situation, the case under consideration is the most
general one.

\

So, let  $\p_0\in\QQQ$. Consider an exact representation of
algebra  \AAA. There exists a vector  $|\Phi_0\rangle$ in the
Hilbert space of this representation such that
$\p_0|\Phi_0\rangle=|\Phi_0\rangle, \quad
\langle\Phi_0|\Phi_0\rangle=1, \quad
\p_0=|\Phi_0\rangle\langle\Phi_0|$. For arbitrary $\B\in\AAA$,
consider the combination $\p_0\B\p_0= |\Phi_0
\rangle\langle\Phi_0|\B|\Phi_0 \rangle\langle\Phi_0|\equiv
\vt(\B)|\Phi_0 \rangle\langle\Phi_0|$, i.e.,
  \beq{23}
\p_0\B\p_0=\vt(\B)\p_0.
  \eeq

Relation~\rr{23} is a relation between elements of algebra \AAA.
Thus, it is a feature of algebra  \AAA{}, and not a property of
the representation.  In particular, functional $\vt(\B)$ is
independent of the representation. It is easily seen that
$\vt(\B)$ is an algebraic state of algebra  \AAA. Its linearity is
implied by the relation
 $$\vt(\B+\D)\p_0=\p_0(\B+\D)\p_0=[\vt(\B)+\vt(\D)]\p_0.$$

Its positivity is a consequence of the relation
$\p_0\B^*\B\p_0=\vt(\B^*\B)\p_0$. Since operators $\p_0\B^*\B\p_0$
and $\p_0$ are positive,  $\vt(\B^*\B)\geq0$. Finally,
normalization is implied by the relation
$\vt(\I)\p_0=\p_0\I\p_0=\p_0$. In addition, restriction of
functional $\vt(\cdot)$ onto the subalgebra  $\QQQ$ is a character
of this subalgebra. Indeed, let $\A\in\QQQ$ è $\B\in\QQQ$. Then
 $$\vt(\A\B)\p_0=\p_0\A\B\p_0=\p_0\A\p_0\p_0\B\p_0=
 \vt(\A)\vt(\B)\p_0.$$

We conclude that functional  $\vt(\cdot)$ has all the properties
required in Postulate 7 from a quantum average. Apart from this,
$\vt(\cdot)$ is positive and satisfies the normalization
condition. These are precisely the conditions that any functional
describing a quantum state should satisfy. Equality  \rr{23} ) is
of a purely algebraic nature. Therefore, the value $\vt(\B)$
depends only on  $\p_0$ (the quantum state) and on  $\B$
considered as an element of algebra  \AAA. However, it does not
depend on any particular commutative subalgebra ($\B$ may belong
to a number of such subalgebras). This means that functional
$\vt(\cdot)$ satisfies Postulate 6.

Let us demonstrate now that the opposite statement is also valid.
If a functional $\Psi^0_{\xi}(\cdot)$  corresponds to a quantum
state $\Psi^0_{\xi}$ that satisfies the condition   $\vx(\p_0)=1$,
then $\Psi^0_{\xi}(\cdot)=\vt(\cdot)$. Indeed, equality~\rr{10}
implies that
  \beq{24}
\Psi^0_{\xi}(\p_0)=\Psi^0_{\xi}(\I)=1.
  \eeq

With the Cauchy-Schwarz inequality (see formula~\rr{18}), we
obtain
 $$\lt|\Psi^0_{\xi}\lt(\B(\I-\p_0)\rt)\rt|^2\leq
\Psi^0_{\xi}B^*\B)\Psi^0_{\xi}(\I-\p_0).$$

From this, in view of equality~\rr{24}, we have
   \beq{25}
\Psi^0_{\xi}(\B)=\Psi^0_{\xi}(\B\p_0)=\Psi^0_{\xi}(\p_0\B).
  \eeq

Substitution $\B\to(\I-\p_0)\B$ in \rr{25} yields
     \beq{26}
\Psi^0_{\xi}(\B)=\Psi^0_{\xi}(\p_0\B\p_0).
  \eeq
Using~\rr{23} in the right-hand-side of~\rr{26}, we obtain the
equality
 \beq{27}
\Psi^0_{\xi}(\B)=\Psi^0_{\xi}\lt(\vt(\B)\p_0\rt)=\vt(\B).
  \eeq

Consider now a GNS construction that uses $\Psi^0_{\xi}(\B)$ as
its generating functional. Let  $\Phi_0(\I)$ be the equivalence
class of the element  $\I$. Then, due to Eqs.~\rr{20} and \rr{22},
we have
 \beq{28}
 \lt(\Phi_0(\I),\Pi(\B)\Phi_0(\I)\rt)= \Psi^0_{\xi}(\B)
  \eeq
for any  $\B\in\AAA$.

According to equalities~\rr{10} and \rr{11},  functional
$\Psi^0_{\xi}(\B)$ describes the average value of the observable
$\B$ in the quantum state  $\Psi^0_{\xi}$. Equality~\rr{28} tells
that this average coincides with the expectation value of the
operator $\Pi(\B)$ over the state described by vector $\Phi_0(\I)$
of the Hilbert space. This observation allows one to employ to
full extent the mathematical formalism of the standard quantum
mechanics for computing quantum averages.

At the same time, there is a considerable difference between the
present approach and standard quantum mechanics. In the latter,
relations similar to~\rr{28} are postulated (the Born
postulate~\cc{born}). This postulate suffices for computations in
quantum mechanics, but its necessity remains unclear. In contrast,
in our approach, equality~\rr{28} is a consequence of
phenomenologically necessary postulates.

Let us pass on to the case when the subalgebra  $\QQQ$ does not
contain one-dimensional projectors. In this case, an exact
representation of algebra  \AAA{} should be considered. Let
$\HHH$ be the Hilbert space of this representation and
$\BB(\HHH)$ be the set of all bounded linear operators in  $\HHH$.
Algebras \AAA{} and $\QQQ$ may be considered as subalgebras of
algebra $\BB(\HHH)$.

Let $\QQ'_{\xi}$ be the maximal real commutative subalgebra of
algebra  $\BB(\HHH)$ such that  $\QQ'_{\xi}\supset\QQQ$. Consider
the set of all projectors belonging  $\QQ'_{\xi}$. These
projectors are mutually commuting self-adjoint operators in $\HHH$
having discrete spectra. There exists an orthonormal basis in
$\HHH$ consisting of the eigenvectors of these operators. Let
$\{\p\}$ be the set of projectors onto such basis vectors. All
these projectors are one-dimensional. They belong to  $\BB(\HHH)$,
but do not belong in the case under consideration to  $\QQQ$. Each
of the projectors  $\p_i\in\{\p\}$ defines a linear functional
$\vt_i(\cdot)$ íà $\BB(\HHH)$: $\p_i\A\p_i=\vt_i(\A)\p_i$.
Restriction of this functional onto the algebra \AAA{}  has all
the properties required for description of the corresponding pure
quantum state.

\section{ILLUSTRATIONS}

Let us consider three simple examples illustrating the general
considerations.

The first example is that of a two-level quantum system, whose
observables are described with Hermitian matrixes $2\times 2$. The
algebra  \AAA{} of dynamical quantities is in this case the set of
all matrixes of the form

$$ \A=\lt[\begin{array}{cc}
  a & b \\
  c & d
\end{array}\rt], $$
with the algebraic operations coinciding with the corresponding
matrix operations.

It is easy to construct all the elementary states for such a
system. Let $\A$ be a Hermitian matrix, i.e.,  $a^*=a$, $d^*=d$,
$c=b^*$. Any such matrix is representable in the form
 \beq{29}
 \A=r_0\I+r\,\hat\tau(\xxi).
 \eeq
Here,  $\xxi$ is a unit three-dimensional vector and  $\tau_i$ are
the Pauli matrixes, $\hat\tau(\xxi)=(\ttau\cdot\xxi)$. Equation
\rr{29} holds if
 \bea{30}
& r=\lt((a-d)^2/4+b\,b^*\rt)^{1/2}, \quad
 r_0=(a+d)/2,\nn & \xi_1=(b+b^*)/(2r), \quad
 \xi_2=(b-b^*)/(2ir), \quad \xi_3=(a-d)/(2r).
 \eea

Evidently, $\hat\tau(-\xxi)=-\hat\tau(\xxi)$. If
$\xxi'\ne\pm\xxi$, the commutator of the matrixes $\hat\tau(\xxi)$
and $\hat\tau(\xxi')$ is nonvanishing. Thus, each $\hat\tau(\xxi)$
(up to a sign) is a generating element of a real maximal
commutative subalgebra  $\QQ_{\xxi}$. Since
$\hat\tau(\xxi)\hat\tau(\xxi)=\I$, the spectrum of element
$\hat\tau(\xxi)$ consists of two points,  $\pm1$.

Let $\vp^{\al}=[\vp^{\al}_{\xxi}]$ be an elementary state. Here,
$\vp^{\al}_{\xxi}$ is a character of subalgebra  $\QQ_{\xxi}$, and
the superscript $\al$ marks different elementary states. Consider
a function  $f^{\al}(\xxi)$ satisfying the restriction
$f^{\al}(-\xxi)=-f^{\al}(\xxi)$ and, for each  $\xxi$, the value
of the function is either  +1, or -1; superscript $\al$ marks
different functions. Evidently, we can  put
$\vp^{\al}_{\xxi}(\hat\tau(\xxi))=f^{\al}(\xxi)$. Taking into
account that $\hat\tau(\xxi)$ is a generating element of
subalgebra $\QQ_{\xxi}$, we obtain that

 $$\vp^{\al}_{\xxi}(\A)=r_0(\A)+r(\A)f^{\al}(\xxi)$$
for any $\A\in\QQ_{\xxi}$.

We point out that any observable  $\A$ (which is not a multiple
$\I$) belongs for this quantum system to one and only one maximal
subalgebra $\QQ_{\xxi}$. Generally, this property certainly does
not hold. Due to this peculiarity of the system under
consideration, we can represent any elementary state as a whole
functional defined on the whole set  $\AAA_+$:
 \beq{31}
\vp^{\al}(\A)=r_0(\A)+r(\A)f^{\al}(\xxi(\A))
 \eeq
for any observable  $\A$. Here,  $r$, $r_0$, and $\xxi$ should be
considered as functions of  $\A$ (see formula~\rr{30}).This
functional can be naturally extended onto the entire algebra \AAA.
This example gives extra evidence against the possibility of
extending the proof of von Neumann for the inexistence of hidden
parameters to elementary states.

Observe that functional $\vp^{\al}(\A)$ defined by formula~\rr{31}
is nonlinear. Generally, an elementary state can be formally
represented as a nonlinear functional defined on the whole algebra
\AAA. However, in the general case, this functional is not single
valued, because one and the same observable $\A$ may
simultaneously belong to several maximal commutative subalgebras
$\QQQ$.

Let us come back to the two-level system. The Hamiltonian of this
system can be represented as
 $$\HH=\lt[ \begin{array}{cc}
  E_0 & 0 \\
  0 & -E_0
\end{array}\rt]$$
and, the projector onto the ground state, as
 $$\p_0=\lt[ \begin{array}{cc}
  0 & 0 \\
  0 & 1
\end{array}\rt].$$

Evidently,  $\p_0\HH\p_0=-E_0\p_0$,
$\p_0\,\po(\A)=\p_0\,\A\,\p_0=\p_0\,d(\A)$. Thus, the ground state
is the linear functional $\po(\A)=d(\A)$. On the other hand, the
ground state is the equivalence class of the elementary states
determined by the condition $\vp^{\al=0}(\p_0)=1$. From this
condition we obtain that
 \beq{32}
f^0(\xi_1=0,\xi_2=0,\xi_3=1)=-1.
 \eeq

We conclude that the ground state is the set of elementary
states~\rr{31} that have function $f^0$ satisfying
condition~\rr{32}.

It is interesting to look at the appearance of the ergodicity
property. We assume that the time evolution of the observables is
defined in the usual way with the unitary automorphism
 $$\A\stackrel{t}{\longrightarrow}\A(t)=\exp(-i\HH t)\A\exp(i\HH
 t)$$
and consider the observable  $\Aa$,
 $$
 \Aa=\lim_{L\to\infty}\frac{1}{2L}\int^L_{-L}dt\,\A(t).
 $$

The limit and the integral are understood in the sense of the weak
topology (see (D.33)). The spectral decomposition
$\HH=E_0\p_1-E_0\p_0$  is valid for $\HH$, where $\p_1=\I-\p_0$.
Thus, observable $\Aa$ can be represented as follows:
 $$\Aa=\lim_{L\to\infty}\frac{1}{2L}\int^L_{-L}dt\,
 \lt[\p_0\A\p_0+\p_1\A\p_1+\p_0\A\p_1 e^{2iE_0t}+
 \p_1\A\p_0 e^{-2iE_0t}\rt]=\p_0\A\p_0+\p_1\A\p_1. $$

Let $\vp^0$ be an elementary state belonging to the ground quantum
state. Then,
 $$\vp^0(\p_1)=\vp^0(\I-\p_0)=0.$$
 Since $[\p_0\A\p_0,\p_1\A\p_1]=0$,
 $$\vp^0(\Aa)=\vp^0(\p_0\A\p_0) +\vp^0(\p_1\A\p_1) =
 \po(\A)+\vp^0(\p_1)\vp^0(\p_1\A\p_1).$$
This relation implies that
 $$ \vp^0(\Aa)=\po(\A).$$

We conclude that the value of the observable $\A$ averaged over
the ensemble described by the ground quantum state equals the
value this observable averaged over time in any elementary state
belonging to the ground quantum state.

Our second example is the one-dimensional harmonic oscillator. We
are interested in the Green's functions of this system. Surely, it
is possible to pass on to the standard scheme that uses the
Hilbert space via the GNS construction. However, it is possible to
suggest a method that is more straightforward from the standpoint
of our approach.

We take that a harmonic oscillator is a physical system described
with an algebra \AAA. of dynamical quantities. This is the algebra
with two Hermitian generating elements $\X$ and $\K$ satisfying
the commutation relation
  $$
  [\X,\K]=i.
  $$

We use the natural units where  $\hbar=m=1$. The time evolution in
algebra  \AAA{} is governed with the Hamiltonian
$\HH=1/2(\K^2+\om^2\X^2)$. Elements $\X$, $\K$, and $\HH$ are
unbounded, and, therefore, they do not belong to $C^*$-algebra.
However, it is possible to use at this point a procedure
frequently employed in the algebraic approach. The procedure
prescribes to consider these elements as given by their spectral
expansions over projectors. The projectors belong on the one hand
to  $C^*$-algebra, and, on the other hand, they define the
representation of these elements as linear operators in the
Hilbert space. The elements admitting such a procedure are called
adjoint to  $C^*$-algebra. Thus, in this case, \AAA{} is the
$C^*$-algebra supplemented with the adjoint elements.

It is convenient to pass on from the Hermitian elements  $\X$ and
$\K$ to the elements $$\am=\frac{1}{\sqrt{2\om}}(\om\X+i\K),
\qquad \ap=\frac{1}{\sqrt{2\om}}(\om\X-i\K)$$ satisfying the
commutation relation
 \beq{33}
 [\am,\ap]=1
 \eeq
and having simple time dependence
 $$\am(t)=\am\exp(-i\om t),
\qquad \ap(t)=\ap\exp(+i\om t).$$

Let us evaluate the generating functional of the Green's function.
In the standard quantum mechanics, the $n$-time Green's function
is defined by the formula
   $$
G(t_1,\dots t_n)=\langle 0|T(\X(t_1)\dots \X(t_n))|0\rangle,
  $$
where $T$ is the time ordering operator, and  $|0\rangle$ is the
quantum ground state.

According to Eqs.  \rr{23} and \rr{27} the Green's function is
defined in our approach with the formula
 \beq{35}
\p_0 T(\X(t_1)\dots \X(t_n))\p_0=G(t_1,\dots t_n)\p_0,
 \eeq
where $\p_0$ is the spectral projector of $\HH$ corresponding to
the minimal value of energy.

It is easily seen that $\p_0$ is representable in the form
 \beq{36}
\p_0 =\lim_{r\to \infty} \exp(-r\ap\am).
 \eeq

As mentioned above, the limit should be understood in the sense of
the weak topology of $C^*$-algebra.

Let us first prove an auxiliary relation:
 \beq{37}
\E =\lim_{r_1,r_2\to \infty}
\exp(-r_1\ap\am)(\ap)^k(\am)^l\exp(-r_2\ap\am)=0 \qquad k,l\ge0,
\quad k+l>0.
 \eeq
 Let $\Psi$ be a bounded positive linear functional. Then,
  $$
\Psi(\E) =\lim_{r_1,r_2\to \infty}\exp(-r_1k-r_2l)\Psi(
(\ap)^k\exp(-r_1\ap\am)\exp(-r_2\ap\am)(\am)^l).
 $$
Here, we used the continuity of the functional $\Psi$ and
commutation relation~\rr{33}. Next, taking into account
$\|\exp(-r\ap\am)\|\le 1$, we obtain
  \begin {eqnarray*}
  |\Psi(\E)|&\leq& \lim_{r_1,r_2\to \infty}\exp(-r_1k-r_2l)
|\Psi( (\ap)^k\exp(-2r_1\ap\am)(\am)^k)|^{1/2}\nn &\times&\quad
|\Psi((\ap)^l\exp(-2r_2\ap\am)(\am)^l)|^{1/2} \nn &\leq&
\lim_{r_1,r_2\to \infty}\exp(-r_1k-r_2l)
|\Psi((\ap)^k(\am)^k)|^{1/2} |\Psi((\ap)^l(\am)^l)|^{1/2}=0
 \end {eqnarray*}

This proves equality  \rr{37}.

Now we can check equality~\rr{36}. In terms of the elements $\ap$,
$\am$, the Hamiltonian $\HH$ is represented as
$\HH=\om(\ap\am+1/2)$. By~\rr{37}
  $$
\lim_{r_1,r_2\to \infty} \exp(-r_1\ap\am)\HH\exp(-r_2\ap\am)=
\frac{\om}{2}\lim_{r_1,r_2\to \infty} \exp(-(r_1+r_2)\ap\am).
   $$

This proves equality ~\rr{36}.

Formula ~\rr{35} implies that
  \beq{38}
G(t_1,\dots
t_n)\p_0=\lt.\lt(\frac{1}{i}\rt)^n\frac{\delta^n}{\delta
j(t_1)\dots\delta j(t_n)} \p_0 T\exp\lt( i\int^\infty_{-\infty}
dt\,j(t)\X(t)\rt)\p_0\rt|_{j=0}.
   \eeq
By Wick's theorem (see ~\cc{bog})
   \bea{39}
&T\exp\lt( i\int^\infty_{-\infty} dt\,j(t)\X(t)\rt)=\nn& =
\exp\lt( \frac{1}{2i}\int^\infty_{-\infty} dt_1dt_2\,
\frac{\delta}{\delta\X(t_1)}D^c(t_1-t_2)\frac{\delta}{\delta\X(t_2)}\rt)
:\exp\lt( i\int^\infty_{-\infty} dt\,j(t)\X(t)\rt):.
 \eea
Here $:\;:$ is the normal ordering operation, and
 $$ D^c(t_1-t_2)=\frac{2}{\pi}\int dE\,\exp\lt(-i(t_1-t_2)E\rt)
 \frac{1}{\om^2-E^2-i0}. $$

Making variations in $\X$ on the right-hand-side of~\rr{39} and
taking into account~\rr{37}, we obtain
   \begin{eqnarray*}
\p_0T\exp\lt( i\int^\infty_{-\infty} dt\,j(t)\X(t)\rt)\p_0&=&
\exp\lt(- \frac{1}{2i}\int^\infty_{-\infty} dt_1dt_2\,
j(t_1)D^c(t_1-t_2)j(t_2)\rt)\nn \times \p_0:\exp\lt(
i\int^\infty_{-\infty} dt\,j(t)\X(t)\rt):\p_0 &=& \p_0 \exp\lt(-
\frac{1}{2i}\int^\infty_{-\infty} dt_1dt_2\,
j(t_1)D^c(t_1-t_2)j(t_2)\rt).
 \end{eqnarray*}

Comparison with~\rr{38} gives
 $$ G(t_1\dots t_n)=\lt.\lt(\frac{1}{i}\rt)^n\frac{\delta^n Z(j)}
 {\delta j(t_1) \dots \delta j(t_n)}\rt|_{j=0}, $$
where
 $$ Z(j)=  \exp\lt(\frac{i}{2}\int^\infty_{-\infty} dt_1dt_2\,
j(t_1)D^c(t_1-t_2)j(t_2)\rt) $$ is the generating functional.

As is known, consideration of field theory models within
perturbation theory can be reduced to consideration of a
multidimensional harmonic oscillator. For this reason, the above
method for computing the generating functional of Green's
functions can be directly generalized to quantum field theory
models.

Our third example is the scattering of a quantum particle off two
slits.

For simplicity, we consider two identical slits, $a$ and $b$. A
homogeneous beam of particles is incident perpendicularly to the
screen with the slits. We are interested in the interference
pattern. Evidently, the structure of the pattern is determined by
the probability distribution of the particle's momenta after the
scattering.

Let $P(F_k)$ be the probability that a particle after scattering
has momentum $K$. More accurately, that it has a momentum inside a
small interval around $F_k$.  $P(F_k)$  denotes the corresponding
event in the sense of probability theory. Probability  $F_k$ is
conditional: before $F_k$ happens, the particle should come
through one of the slits. This means that the coordinate of the
particle at the moment of scattering should be either near $a$ (we
denote this event  $F_a$) or near $b$ (event $F_b$). The
conditional probability is then denoted as $P(F_k/(F_a+F_b))$.

By the probability theory (see, e.g.,  \cc{nev}):
 \beq{391}
 P(F_k/(F_a+F_b))=\frac{P(F_k\cap(F_a+F_b))}{P(F_a+F_b)}.
 \eeq

$F_k\cap(F_a+F_b)$ means that two events have happened: $F_k$ and
$(F_a+F_b)$. The symbol $\cap$ is used in order to take into
account that the subset of elementary events corresponding to two
simultaneous events coincides with the intersection of the subsets
of elementary events corresponding to each of the events.

Since the events $F_a$ and $F_b$ do not intersect, we have
 \beq{392}
 P(F_k\cap(F_a+F_b))=P(F_k\cap F_a+F_k\cap F_b)
  =P(F_k\cap F_a)+P(F_k\cap F_b).
  \eeq

On the other hand, because the beam is homogeneous,
  \beq{393}
 P(F_a+F_b)=P(F_a)+P(F_b)=2P(F_a)=2P(F_b).
    \eeq
With formulas \rr{392} and \rr{393}, equality~\rr{391} can be
transformed to
 $$P(F_k/(F_a+F_b))=\frac{1}{2}\lt[\frac{P(F_k\cap F_a)}{P(F_a)}
 +\frac{P(F_k\cap F_b)}{P(F_b)}\rt].$$

The first term in the square bracket of the right-hand side comes
from scattering off the slit  $a$ and the second from scattering
off the slit $b$. We did not obtain any interference. The mistake
was that we used the formulas of probability theory without taking
into account the specifics of applying probability theory to
quantum events. Formula~\rr{391} is already erroneous. The problem
is that the momentum and coordinate observables are incompatible.
Therefore, as we explained in section 4, no probability can be
accorded to subset $F_k\cap(F_a+F_b)$.

A way to avoid this mistake is to take that the first stage of
scattering consisting in the appearance of a particle near one of
the slits is a preparation of a quantum state, which has its
equivalence class of elementary states. Next, this class should be
considered as the new space of elementary events. In this space,
scattering at a given angle can be considered as an unconditional
event.

We relate with event $F_a$ (particle appeared in domain near slit
$a$) an observable $\p_a$ that can take two values, unity if
particle passed through the slit, and zero otherwise. Evidently,
this observable has the properties of a projector. Analogously, we
introduce an observable $\p_b$. Clearly, the interference pattern
is formed by particles whose elementary states correspond to the
unit value of the observable $\p_a+\p_b$. We denote the set of
these elementary state as $\Om(p_a+p_b=1)$. This set corresponds
to a quantum ensemble. Generally, such an ensemble is a mix of
pure ensembles. The average values of observables over each of
these pure ensembles are described by a linear positive
functional. Thus, the average values of the observables over the
quantum ensemble under consideration are also described by a
linear positive functional.

Let $\Psi(\cdot)$ be such a functional corresponding to the set
$\Om(p_a+p_b=1)$. This functional satisfies the equality .
 $$\Psi(\p_a+\p_b)=\Psi(\I)=1.$$
Thus, we can reproduce the derivation of formula~\rr{26} for this
functional, and justify the equality
 $$\Psi(\B)=\Psi\lt((\p_a+\p_b)\B(\p_a+\p_b)\rt)$$
for any observable $\B$. For the average value of the momentum,
this formula yields
 $$\langle\K\rangle=\Psi(\p_a\K\p_a)+\Psi(\p_b\K\p_b)+
 \Psi(\p_a\K\p_b+\p_b\K\p_a).$$

The first two terms in the right-hand side are the contributions
from the scattering off slits $a$ and $b$, respectively. The third
term is the interference term.

\section {PROBLEM OF PHYSICAL REALITY}

Einstein, Podolsky, and Rosen \cc{epr} formulated the necessary
principles for constructing a complete physical theory: (a) "every
element of the physical reality must have a counterpart in the
physical theory" and (b) "if, without in any way disturbing a
system, we can predict with certainty (i.e., with probability
equal to unity) the value of a physical quantity, then there
exists an element of physical reality corresponding to this
physical quantity."

Standard quantum mechanics does not comply with these theses. A
single experiment has no copy in the mathematical formalism of
standard quantum mechanics. Moreover, there is a strong opinion
that such a copy cannot exist, and that even the objective
physical reality that determines the outcome of a single physical
experiment does not exist.

This opinion is backed with strong arguments. Probably, the best
known one is based on the Bell inequality~\cc{bel1,bel2}. Bell
derived his inequality within the framework of the EPR theses.
Following Bell, many versions of similar inequalities were
suggested. Here, we consider the version suggested in~\cc{chsh}.
This version is usually denoted as CHSH.

Let a particle of zero spin decay into two particles $A$ and $B$,
each having spin 1/2. These particles move away from one another
by large distances and are detected by devices $D_a$ and $D_b$,
respectively. Device $D_a$ measures spin projection onto direction
$a$ for particle $A$, $D_b$ measures spin projection onto
direction $b$ for particle $B$. The corresponding observables are
denoted as $\A_a$ and $\B_b$, and the outcomes of the measurements
as $A_a$ and $B_b$.

Let us assume that the state of the initial particle is
characterized with a physical reality, which is parameterized with
parameter $\nu$. The same parameter is used to describe the
physical realities that characterize the decay products.
Correspondingly, the  outcomes  of measurements of observables
$\A_a$ and $\B_b$ are functions of the parameter $\nu$, $A_a(\nu)$
and $B_b(\nu)$. Let the distribution of events in parameter $\nu$
be characterized with the probability measure $P(\nu)$ satisfying
the standard conditions:
 $$ \int P(d\nu)=1, \qquad 0\leq P(\nu)\leq 1. $$

 Let us introduce the correlation function   $E (a, b) $:
\beq{394}
  E(a,b) = \int P(d\nu)\,A_a(\nu)\,B_b(\nu)
\eeq
 and consider the combination
  \bea {40}
N&=&|E (a, b) -E (a, b ') | + |E (a ', b) +E (a ', b ') | = \nn{}
 &=& \lt |\int P(d\nu)\,
A_a(\nu)\,[B_b(\nu)-B_{b'}(\nu)]\rt|+ \lt |\int P(d\nu)\,
A_{a'}(\nu)\, [B_b(\nu) +B_{b'}(\nu)]\rt|.
  \eea
For any directions $a$ and $b$,
 \beq {41}
 A_a (\nu) = \pm1/2, \quad B_b (\nu) = \pm1/2.
  \eeq

 Thus,
\bea{42}
 N & \le &\int P(d\nu)\,[|A_a(\nu)|\,|B_b(\nu)-B_{b'}(\nu)|+ |A_{a '}(\nu)
| \, |B_b (\nu) +B_{b '} (\nu) |] =\nn{} &=&1/2 \int P(d\nu) \,
[|B_b(\nu)-B_{b'}(\nu)|+|B_b(\nu)+B_{b'}(\nu)|].\eea

Due to equalities~\rr{41}, one of the two expressions
 \beq{43}
 |B_b(\nu) -B_{b'} (\nu) |, \qquad |B_b (\nu) + B_{b'}(\nu) | \eeq
vanishes, while another equals unity. We point out that both
expressions involve the same value of~$\nu$.

With these properties of expressions~\rr{43}, inequality~\rr{42}
implies the Bell inequality (CHSH):
 \beq{44} N \leq 1/2\int P(d\nu) =1/2. \eeq

Within the standard quantum mechanics, the correlation function is
easily calculated. The result is as follows:
 $$
 E (a, b) = -1/4\cos\theta_{ab},
 $$
where $ \theta_{ab}$ is the angle between the directions  $a$ and
$b$. For the directions  $a=0$, $b=\pi/8$, $a'=\pi/4$,
$b'=3\pi/8$, we obtain
 $$ N=1/\sqrt{2}, $$
which is in contradiction with inequality~\rr{44}.

Experiments agree with the quantum mechanical computations and
disagree with the Bell inequality. Usually, these results are
considered as a manifestation of the fact that no physical reality
corresponding to a quantum mechanical system would predetermine
the results of a measurement.

However, from the standpoint of modern probability theory, the
above derivation of the Bell inequality is too naive. This
derivation assumes that there is a probability distribution in the
parameter $\nu$. By its meaning, this parameter marks an
elementary event. As was pointed out (see Section 4), it is not
always possible to assign a probability to an elementary event.
The set of elementary events should be equipped with the structure
of measurable space before one can speak about probability. In
view of this, let us try to reproduce derivation of the Bell
inequality using elementary state $\vp$ in place of the parameter
$\nu$.

According to the problem's conditions, the initial particle is in
a particular quantum state. Thus, the space $\Om(\vp_{\eta})$ of
elementary events $\vp$ is the equivalence $\{\vp\}_{\vp \eta}$.
Thus, if an observable $\A\in\QQ_{\eta}$, then the observable will
take one and the same value on all $\vp\in\{\vp\}_{\vp \eta}$. The
difference between elementary states  $\vp\in\{\vp\}_{\vp \eta}$
can be resolved by the values of observables $\B\notin\QQ_{\eta}$.
It is easily seen that due to this difference the set
$\vp\in\{\vp\}_{\vp \eta}$ have the cardinality of continuum. Let
us consider a subalgebra $\QQQ\neq\QQ_{\eta}$ to justify this.
Since the subalgebras $\QQQ$ and $\QQ_{\eta}$ are maximal, there
exists at least one observabable $\B$ such that  $\B\in\QQQ$  and
$\B\notin\QQ_{\eta}$. The spectrum of this observable cannot be a
single point. If $\lll$ were this single point of the spectrum,
then spectral radius of the element $\B-\lll\I$ would vanish:
$r(\B-\lll\I)=0$. However, $\|\B-\lll\I\|=r(\B-\lll\I)$ for
$C^*-$algebra. Thus,  $\B=\lll\I\in\QQ_{\eta}$. We conclude that
there exist at least two elementary states $\vp\in\{\vp\}_{\vp
\eta}$ giving different values of observable $\B$. The same
reasoning is applicable to any subalgebra  $\QQQ\neq\QQ_{\eta}$.
Since the set of such subalgebras  $\QQQ$ has the cardinality of
continuum, the set of different $\vp\in\{\vp\}_{\vp \eta}$ will
also have the cardinality of continuum.

Consider now formula~\rr{394}) for the correlation function. We
need correlation functions for four combinations of the
observables: $\A_a\B_b$, $\A_a\B_{b'}$, $\A_{a'}\B_b$, and
$\A_{a'}\B_{b'}$. We are interested in the case when the
directions $a$, $a'$, $b$, and $b'$ are not parallel to one
another. In this case, the listed observables are not compatible
with one another. Therefore, obtaining experimentally correlation
functions requires four separate series of experiments. In
reality, each of these series has a finite number of experiments.
Ideally, they each would be a countable set of experiments.

We conclude that in experiment we deal not with a whole space
$\Om(\vp_{\vp\eta})$ of elementary events, but with four random
samples  $\Om_{ab}$, $\Om_{ab'}$, $\Om_{a'b}$, and $\Om_{a'b'}$.
Since these samples are countable even in the ideal case, and the
set $\Om(\vp_{\vp\eta})$ has the cardinality of continuum, the
probability that these samples contain common elements is zero. In
addition, to make these samples measurable, one should select
corresponding $\sss$-algebras  $\F_{ab}$, $\F_{ab'}$, $\F_{a'b}$,
and $\F_{a'b'}$. These subalgebras differ from one another.
Moreover, as discussed in section 4, they cannot be subalgebras of
a single $\sss$-algebra that would have a probability measure
corresponding to it. In other words, each sample should have its
own probability measure: $P_{ab}$, $P_{a'b}$, $P_{ab'}$,
$P_{a'b'}$.

Now formula~\rr{394}) takes the form
 $$ E(a, b)=\int_{\Om_{ab}}P_{ab}(d\vp)\vp(\A_a\B_b), $$
and~\rr{40} becomes
 \begin{eqnarray}\label{45}
N&=& \lt |\int_{\Om_{ab}}P_{ab}(d\vp)\, \vp(\A_a\B_b)-
\int_{\Om_{ab'}} P_{ab'}(d\vp)\,\vp(\A_a\B_{b'})\rt|+ \nn{} & +&
\lt |\int_{\Om_{a'b}} P_{\A_{a'}\B_b}(d\vp)\, \vp(\A_{a'}\B_b) +
\int_{\Om_{a'b'}} P_{a'b'}(d\vp)\, \vp(\A_{a'}\B_{b'})\rt|.
  \end{eqnarray}

The same symbol $d\vp$ is used in all four terms of
formula~\rr{45}. Despite this, the sets of elementary states
corresponding to $d\vp$ are different in different terms. They are
elements of different $\sss$-algebras. Moreover, the probability
of sharing elements between these algebras vanishes. We conclude
that, first, joining integrals under the modulus~\rr{45} is
erroneous (this step was used in obtaining formula~\rr{40}).
Second, forming pairs like the ones featured in formula~\rr{43} is
erroneous. Thus, the proof of inequality~\rr{44}is invalid. Our
conclusion is that if we relate physical reality to the elementary
state, violation of Bell's inequality does not demonstrate in any
way the contradictory nature of this notion.

Another argument against the applicability of the notion of
physical reality to quantum physics is the Kochen-Specker no-go
theorem~\cc{ksp}. The meaning of this theorem can be explained as
follows. Consider as a physical system a particle with unit spin.
Let $x$, $y$, $z$ be mutually orthogonal directions. Then
observables $\s_x^2,\s_y^2,\s_z^2$ (spin projections squared)
commute with each other. Thus, they are compatible and can be
measured simultaneously. Assume that there exists a physical
reality that predetermines unambiguously the outcome of a
measurement for any direction. Measurement along some direction
should yield zero, and along two others, unity. Select one of the
latter directions and consider two directions (different from the
initial directions) that are perpendicular to the selected
direction. One of these directions should yield zero and another
unity. Select the one yielding zero and repeat the whole procedure
from the very beginning. It is possible to get in a finite number
of such steps to a direction previously encountered. If this
happens, it turns out that, if the first value corresponding to
the direction was zero, the value obtained at the second encounter
will be unity.

The conclusion drawn from this contradiction is that the physical
reality governing the outcomes of measurements cannot exist.
However, the above reasoning ignores completely the problem of
measurability. Meanwhile, here we deal with two triples of
directions: $x,y,z$, and  $x,y',z'$. Within each triple, all the
directions are orthogonal, but there are nonorthogonal directions
in different triples. Therefore, observables
$\s_x^2,\s_y^2,\s_z^2$ and $\s_x^2,\s_{y'}^2,\s_{z'}^2$ belong to
different commutative subalgebras of algebra \AAA.
Correspondingly, devices performing measurements that are
compatible within each of the triples belong to different types.
These devices should not necessarily yield the same outcome when
measuring the observable $\s_x^2$. This was tacitly assumed in the
above reasoning. Recall that the elementary state does not fix
unambiguously the values of all observables. It fixes
unambiguously the reading of all the devices of {\it a specific
type}. For different types, these readings may differ. We conclude
that within our approach the Kochen-Specker theorem does not
exclude the existence of a physical reality related to the
elementary state.

\section {PARADOXES}

Critics of standard quantum mechanics have pointed out a number of
situations where quantum mechanical considerations lead to
paradoxes. Here, we discuss only two such paradoxes, selecting the
ones that are probably most frequently mentioned. The first is the
Einstein-Podolsky-Rosen (EPR) paradox and the second, the
Schrodinger cat paradox. It should be noted from the very
beginning that the existence of paradoxes is denied by the most
orthodox partisans of standard quantum mechanics. They claim that
the correct use of the formulas of standard quantum mechanics does
not lead to any paradoxes. So, before discussing specific
paradoxes, we state our own position. It is as follows.

The formulas of standard quantum mechanics are certainly valid
when applied to the description of quantum ensembles. They give
correct values to the observable quantities and event
probabilities also in the physical models considered by the
authors of the paradoxes. Thus, only discussing individual events
is of interest. Here, two positions are possible. On this, see
review~\cc{hom}. First, it is possible to take that individual
events are beyond the consideration of standard quantum mechanics.
This position eliminates the subject of the discussion. However,
individual events do exist. So, the question arises as to the
completeness of the quantum mechanical description. Second, it is
possible to take that quantum mechanics predicts only the
probabilities of individual events, and the completeness of its
description is exhausted by predictions of the probabilities. In
this position, it is necessary to admit that probability is an
independent entity of the individual event.

In modern probability theory, the latter is not the case. Recall
that the notion of the space of elementary events comes before the
notion of probability measure. Correspondingly, an individual
event (elementary event) is considered as an element of a set
(ensemble). And one and the same individual event can be
considered as an element of different sets. Depending on this set,
different probabilities (or no probability) may correspond to one
and the same event.

The orthodox partisans of standard quantum mechanics reject this
point of view and prefer to take that probability is a fundamental
indefinable entity of any individual event, which is represented
in the mathematical formalism of quantum mechanics with either a
vector of a Hilbert space, or a density matrix. Formally,
paradoxes are avoided in this way, but the physical essence of the
phenomena remains beyond the framework of the discussion.

After these preliminary remarks, we begin the discussion of the
paradoxes. We start with EPR. In the original paper~\cc{epr}, the
paradox was considered on the example of measurements of
coordinate and momentum. Âohm suggested a simpler physical
model~\cc{bom1}. The same problem is discussed on the example of
measurements of spin projections onto different directions. Here,
we discuss Bohm's version. In this case, we consider the physical
system that was considered in the discussion of Bell's inequality.

Let a particle of spin 0 decay into two particles, $A$ and $B$, of
spins 1/2. After decay, the particles are separated by a large
distance. According to standard quantum mechanics, the spin state
of this system is described by the state vector
 \beq{46}
 |\Psi\rangle=\frac{1}{ \sqrt{2}}\lt[|A_z^{(+)}\rangle|B_z^{(-)}\rangle
 -|A_z^{(-)}\rangle|B_z^{(+)}\rangle\rt],
 \eeq
where $|A_z^{(\pm)}\rangle$, $|B_z^{(\pm)}\rangle$ are the
eigenvectors of the spin projection operators for axis z with the
eigenvalues +1/2 and -1/2. This is a so-called entangled state. In
this state, neither particle $A$ nor $B$ has any definite value of
the spin projection onto axis $z$. The spin state of each particle
can be described with a density matrix. For example, the density
matrix for particle $A$ is
 $$\rho(A)=\frac{1}{2}\lt[|A_z^{(+)}\rangle\langle A_z^{(+)}| +
 |A_z^{(-)}\rangle\langle A_z^{(-)}|\rt]. $$

This means that the particle has spin projections +1/2 and -1/2
with probabilities 1/2.

Measure projection of spin onto axis $z$ for particle $B$ when two
particles are in space-like domains. Let the result be +1/2. In
this case, by the postulate of instant collapse of the quantum
state (projection principle), the state $|\Psi\rangle$ will be
instantly replaced with the state
 \beq{48}
|\tilde\Psi\rangle=\p_+|\Psi\rangle/
\sqrt{\langle\Psi|\p_+|\Psi\rangle}, \eeq
 where $\p_+$ is a projector of the form
 \beq{49}
\p_+=\I_A\otimes|B_z^{(+)}\rangle\langle B_z^{(+)}|.
 \eeq
Here, $\I_A$ a is the unit operator in the space of states of
particle $A$.

Substituting  \rr{49} into \rr{48}, we obtain
$|\tilde\Psi\rangle=-|A_z^{(-)}\rangle\langle B_z^{(+)}|\rangle$.
The density matrix of particle $A$ corresponding to this state is
$\tilde\rho(A)=|A_z^{(-)}\rangle\langle A_z^{(-}|$. This means
that any subsequent measurement of the axis $z$ spin projection
for particle  $A$ yields the value -1/2 with unit probability.
This is what has been seen in the experiments. Thus, the
projection principle is a good recipe. However, it would be nice
to understand the physical mechanism that provides for the success
of this recipe.

Two variations of such a mechanism are readily available. The
first one is as follows. When created, particles acquire definite
axis $z$ spin projections (of opposite sign). Before measuring
this projection for particle $B$, we do not know the values of the
projections. After the measurement for particle $B$, we know the
projection for particle $A$ automatically. However, this mechanism
contradicts standard quantum mechanics.

This is the case because quantum state $|\Psi\rangle$ can also be
represented in the form
 $$|\Psi\rangle=\frac{1}{\sqrt{2}} \lt[|A_x^{(+)}\rangle|B_x^{(-)}\rangle
 -|A_x^{(-)}\rangle|B_x^{(+)}\rangle\rt],$$
  where the notations are the same as in~\rr{46}, but projections onto
axis $z$ are replaced with ones onto axis $x$. However, the
observables corresponding to spin projections onto axes $z$ and
$x$ are mutually incompatible, and, by the standard quantum
mechanics, they cannot have definite values simultaneously.

The second version of the mechanism is as follows. After the
decay, particles did not acquire any definite value of spin
projection onto any axis. Due to the measurement of the projection
onto a specific axis, they gained definite projections onto this
axis. For particle $B$ that interacted with the measuring device,
this mechanism is viable. However, how could such measurement
influence particle $A$ located in a space-like domain with respect
to the measuring device? This is not possible without a violation
of the principles of relativity theory. Thus, both variants of the
physical mechanism are unsound. This is the paradox.

Objecting against the paradox, Bohr pointed out \cc{bohr3} that it
is not allowed in considering a correlated system to treat it as
consisting of two separate independent parts. And any measurement
acting on a part of the system should be considered as acting on
the whole system. This reasoning is not very convincing in our
opinion. Here, it is important that there are two types of
correlations admitting rational explanations. The first type
originates from interaction between the parts of the system. In
the EPR case, such interaction would have to propagate faster than
light. The second type originates from some constraint on the
initial conditions for the particles under consideration. In the
EPR case, such a constraint is in place since the particles
originate from a decay of a single initial particle. However, the
presence of the constraint is not enough for {\it an unambiguous}
subsequent correlation of the particles. It is also necessary that
the initial conditions would determine uniquely the subsequent
time evolution of the particles. In this case, right after the
creation, before measurement, particles $A$ and $B$ should posses
a property that would determine unambiguously the result of the
measurement. The latter contradicts the concept of standard
quantum mechanics.

Surely, one could assume that there is a special quantum type of
correlation that resists any rational treatment. Such an
assumption would be the worst of all possible ones from the
standpoint of science, since science strives to reduce the number
of truths resisting rational understanding.

More apt are Fock's considerations~\cc{fok2}. Fock had taken that
the notion of state does not have any objective meaning in the
quantum case. Rather, it should be understood as a "data on the
state." With this interpretation, the paradox can be avoided.
However, it raises a question: "Is there anything objective about
which we are collecting data?."

Within the approach of this paper, this "something" does exist. It
is the elementary state. The elementary state is an objective
characteristic of a physical system. It is independent of any
knowledge of the system. In contrast, a quantum state, i.e., some
equivalence class of elementary states is not a completely
objective characteristic of a physical system. Instead, it is an
objective characteristic of an ensemble of physical systems. A
particular system of interest can be considered as an element of
different ensembles (the freedom of choice). Correspondingly, the
system will be characterized with different quantum states. For
this reason, a quantum state involves a subjective factor.

Turning directly to the EPR paradox, we can give it the following
interpretation. Before and after the decay of the original
particle, the physical system has stable (zero) values of the
observables $\s_{\bf n}$ (projections of the total spin onto the
direction ${\bf n}$). After decay, the values of observables
$\A_{\bf n}$ and $\B_{\bf n}$ (spin projections onto direction
${\bf n}$ for particles $A$ and $B$, respectively) satisfy the
relation
 \beq{51}
  A_{\bf n}+B_{\bf n}=S_{\bf n}=0.
  \eeq

In principle, each of the observables $\A_{\bf n}$ and $\B_{\bf
n}$ could be unstable. However, as pointed out in section 6, for a
two-level system (the case of a particle with spin 1/2), these
observables are stable. In the elementary state, incompatible
observables can simultaneously have a definite value. However,
these values cannot be measured simultaneously with any classical
device. In a particular experiment, we can measure the observable
$\B_{\bf n}$ for any but a single direction ${\bf n}$. This is the
case because observables $\B_{\bf n}$ and $\B_{\bf n'}$ are
incompatible for different directions ${\bf n}$ and ${\bf n'}$.
Due to equality~\rr{51}, in such a measurement we automatically
measure $\A_{\bf n}$. This is the so-called indirect measurement.
Thus, in this approach, the paradox is trivially solved.

Using this example, we can rationally treat the quantum state
collapse phenomenon. Within the standard quantum mechanics, this
phenomenon appears mystical.

Before measuring the spin of the  $B$ particle, we know that our
physical system is in the elementary state that belongs to the
equivalence class characterized by the zero values of the
observables $\s_{\bf n}$. In other words, we know that the system
is in a singlet quantum state, but we do not know its elementary
state. After measurement of observable $\B_{\bf n}$, due to
equality~\rr{51}, we gain knowledge not only on the value of this
observable, but also on the value of the observable $\A_{\bf n}$.
Therefore, after the measurement, we know that the system is in
the elementary state that belongs to the equivalence class
characterized by the values $A_{\bf n}=-B_{\bf n}$ ($B_{\bf n}$
known) of the observables $\A_{\bf n}$ and $\B_{\bf n}$. Here, we
assumed that measurement of $\B_{\bf n}$ is reproducible. Now we
again do not know the elementary state of the system, but we know
that the system is in a particular quantum state (of the type
$|\tilde\Psi\rangle$), formula~\rr{48}).

Due to interaction with the measuring device, the value of
observables $\B_{\bf n'}$ for directions ${\bf n'}\neq{\bf n}$
changes uncontrollably. Because of this change, equality~\rr{51}
breaks for such directions. This means that the system ceases to
belong to the singlet state. In this way, all the features of the
collapse of the quantum state are reproduced. We point out that
before the measurement, we could describe the quantum state of
particle $A$ with the density matrix
 \beq{52}
 \rho(A)=\frac{1}{2}\lt[|A^{(+)}_{\bf n}\rangle \langle
A^{(+)}_{\bf n}|+|A^{(-)}_{\bf n}\rangle \langle A^{(-)}_{\bf
n}|\rt] \eeq
 and after the measurement, with the density matrix
 \beq{53}
 \tilde\rho(A)=|-B_{\bf n}\rangle\langle-B_{\bf n}|.
 \eeq

Despite the difference of the quantum states~\rr{52} and~\rr{53}
before and after measurement (the former is mixed, the latter is
pure), we cannot conclude that there is a change in the elementary
state of particle $A$ due to the measurement. We simply gained
extra information on this elementary state.

Equality \rr{51} admits another useful interpretation. In the
decay of the initial particle, each of the secondary particles
measures the elementary state of its partner. This means that the
elementary state of one of the particles is a negative copy of the
elementary state of another particle. The creation of such a copy
can be called a measurement with a quantum device. One particle is
a quantum measuring device for another. In contrast to the
measuring technique using a classical device, such measurement can
fix unambiguously the elementary state of the particle under
measurement. However, gaining access to the results of such a
measurement requires a measurement in which the quantum device is
studied with a classical device. Such a measurement results only
in knowledge of the equivalence class containing the elementary
state of the particle under measurement.

The scenario for the second paradox we will discuss was suggested
by Schrodinger~\cc{sch} (see also~\cc{btk}). It goes as follows. A
cat and a radioactive source of very small intensity are put in a
box. When an atom decays in the source, the event is registered
with a Geiger counter. The pulse from the counter comes through an
amplifier to an automaton that breaks an ampoule with a poison.
The cat is killed by the poison. The observer does not know if the
decay had taken place or not. Thus, by the rules of quantum
mechanics, the observer should describe the state of the complex
system (the cat and the radioactive source) by a state vector that
is a superposition of two quantum states: atom before decay and
live cat plus decayed atom and dead cat. Superposition of the dead
and alive cats looks very odd.

The opinion is sometimes expressed that the paradox disappears if
one goes over from the description in terms of the vectors of a
Hilbert space to description in terms of density matrix. Here, we
should agree on the rules of our game. If we take that the density
matrix describes an ensemble of physical systems, there will be no
paradoxes. In this case, we deal not with a cat, but with an
ensemble of cats, some of which are alive and some are dead. In
this case, each cat is either live or dead, and which one we are
dealing with is governed by probability theory. However, the
meaning of the Schrodinger paradox is that we deal with a single
cat. In this case, the above understanding of the density matrix
is not appropriate. If we take that the density matrix describes
the state of a single cat, a mixed state of a live and dead cat is
not easier to contemplate than the superposition of such cats.

Certainly, there is no paradox if we take Fock's interpretation,
i.e., if we take that the quantum mechanical state contains our
knowledge of the objective state of a physical object. However,
standard quantum mechanics does not take this attitude, and, in
any case, it is not known if the objective state exists.

Within the framework of the elementary state concept, the paradox
is trivially solved. The pair under study (the cat plus the
radioactive atom) is in a specific elementary state. At a
particular moment of time, the cat is either alive or dead in this
state. There is no mixed elementary state of a live and dead cat.
The quantum state describes an equivalence class of such
elementary states. Among these elementary states, there are such
that correspond to a cat alive at the given time, and there are
such that correspond to the cat dead at the same moment of time.

When we place cat in the box, the information available to us
pinpoints only the equivalence class, and not  the elementary
state. The equivalence class is fixed by the conditions we can
classically register: at the moment the system was prepared, the
cat was alive, and the atom had not decayed. On the other hand,
the unambiguous evolution of the system is fixed namely by its
elementary state. This state cannot be fixed unambiguously by
classical measurements.

\section {FIELD-PARTICLE DUALITY}

In quantum case, elementary state $\vp=[\vx]$ of an individual
physical system is a collection of functionals $\vx(\cdot)$, each
of which is a character of a maximal real commutative subalgebra
$\QQQ$ of algebra \AAA. The set $\Xi$ ($\xi\in\Xi$) of these
subalgebras has the cardinality of continuum. Thus, the elementary
state is a field over the set $\Xi$ whose values are functionals.

Setting $\vp$ is equivalent to setting $\vx$ for every
$\xi\in\Xi$. In turn, it suffices to set the value of $\vx$ for
every element of the generating set of the subalgebra $\QQQ$ for
setting $\vx$. It is possible to take the view that a component of
this functional-valued field corresponds to every generating set
of the subalgebra $\QQQ$. The value of functional $\vx$ on a
generating element can be considered as a value of the component
of the field $\vp$ at the point $\xi$. Thus, $\vp$ is a real
$c$-number multicomponent field over the set $\Xi$.

Correspondingly, even a quantum system that is conventionally
considered as a system of finite number of degrees of freedom
(e.g., harmonic oscillator), is, in fact, a field system, i.e., a
system with an infinite number of degrees of freedom. This implies
that the elementary state of any quantum system may principally
encode an infinite amount of information. However, it is not
possible to really use this infinite volume of information. This
is the case because we should have the possibility to control the
information with classical devices to make it usable. However,
classical devices are unable to tell the difference between
different elementary states; they can tell the difference only
between the equivalence classes corresponding to the quantum
states. Thus, the volume of controllable information turns out to
be finite. Still, it can be much larger than for classical
physical systems. This is the physical prerequisite for the
possibility of constructing quantum computers.

The elementary state of any physical system is also a field over
Minkowski space. This is the case because the systems that are
traditionally considered in quantum mechanics as pointlike are in
fact distributed over Minkowski space.

For example, consider an electron that scatters "elastically" off
a nucleus. In reality, such scattering is accompanied by
bremsstrahlung of soft photons. However, the energy and other
observables of these photons are beyond the sensitivity of the
measuring devices. In other words, electron scattering is
accompanied by the emission of soft photons. Although this field
is not mea- surable with classical devices, it plays an important
role in theory. Without it the theory suffers from infrared
divergences. At one time, this situation was called the infrared
catastrophe.

The infrared catastrophe can be avoided only if the number of soft
radiated photons is infinite. This means that the electron is
accompanied by an effective classical field. Clearly, any other
process involving the creation of virtual electron-positron pairs
will be accompanied by such a classical electromagnetic field
(effective). A similar situation takes place in quantum
chromodynamics. In chromodynamics, the major hopes for explaining
quark confinement are related precisely to soft partons.

We conclude that elementary state $\vp$ of any quantum system is
described with a $c$-number field over the set $\Xi$ and Minkowski
space. This field has all the attributes of a real classical
field. If we assume the classical paradigm, we should take that
there exists a material field mathematically accounted for by
$\vp$. This effective classical field can include the classical
component of the electromagnetic field, the gravitation field, or
some other field. Concrete physical realization of this field is
not important for us. We will call it the phase field, by analogy
with the phase space that determines state of the system. We point
out that the field is classical and, therefore, the quantum
relation between energy and frequency may not hold.

Assumption of the material existence of a phase field may help to
solve one of the problems of quantum theory, the problem of
wave-particle duality. Here, we call this notion field-particle
duality. The field properties of a quantum system are naturally
related to the phase field, i.e., to the elementary state.

The particle properties of a quantum system mean the following.
The physical system has local observables, i.e., observables
associated with a bounded domain of Minkowski space. These
observables or, more accurately, their complex combinations form
the algebra of local observables. Note that the algebra of local
observables is one of the major notions of the traditional
algebraic approach to quantum field theory
(see~\cc{araki,haag,emch,hor}). There exist stable (meaning
frequently encountered) sets of values of local observables, which
we treat as quantum particles of particular species: electrons,
photons, nuclei, atoms, etc.

Measuring devices perceive these observables as an indivisible
whole. In this, the particle properties of quantum systems are
exhibited. The reaction of a measuring device is determined by the
elementary state of the system (by the phase field). In turn, the
structure of (the value taken by) the phase field is determined by
the spectra of the corresponding observables. We point out that a
point of the spectrum is an indivisible whole. In this way, the
particle and field properties turn out to be tightly intertwined
in a quantum system.

In standard quantum mechanics, the quantum state of a physical
system is also associated with a $c$-number field, with the wave
function. However, the wave function takes values in complex
numbers. So, it cannot correspond directly to a material field. In
our approach, the wave function is related only to probability,
and the relation is rather indirect. Namely, the average values of
the observables may be represented in the form of expectation
values of linear operators in a Hilbert space (the averages by
their nature are not such objects). In turn, the vectors of
Hilbert space may be represented as wave functions.

The phase field may be considered as a material substance bearing
information on the physical state of a quantum object. It should
be coordinated (coherent) with the associated quantum object. If
this condition is fulfilled, it is possible to construct a
plausible model of the measurement process based on the phase
field. Recall that the lack of such a model is usually put forward
as the reason behind all the hitches that are present in standard
quantum mechanics.

Let us describe such a model. (See also on this subject the papers
of Blokhintsev~\cc{blo1,blo2,blo3,blo4}].) The measuring device
consists of an analyzer and detector. Sometimes these components
are combined. The analyzer is a device with a single input channel
and a number of output channels. If a device is dedicated to
measurement of an observable $\A$ , each output channel
corresponds to a particular region of the spectrum of the
observable, i.e., each output channel corresponds to an
equivalence class of elementary states.

The phase field associated with the object under measurement
excites collective oscillations in the device that are coherent
with the field. The oscillations may be very weak, but due to the
coherence they interact with the quantum object in a resonance
way. Microscopic description of such interaction is practically
inaccessible. However, the result of this interaction can be
described with a boundary condition. If the quantum state of the
object under measurement descries the equivalence class
corresponding to an output channel, the object unavoidably ends up
in this output channel. If the object under measurement is in a
quantum state that does not correspond to any of the output
channels, the analyzer turns out to be a bifurcation domain for
the object. In this case, the resonance interaction of the object
with the oscillations of the analyzer excited by the phase field
is the random force directing the object into a particular output
channel, namely, into the channel corresponding to the equivalence
class that includes the elementary state of the object under
measurement.

Here, the localization domain of the quantum object is the domain
of localization of its local observables that can be registered by
classical measuring devices. In the following, this localization
domain is called the kern of the quantum object. At the same time,
as mentioned earlier, any quantum object is accompanied by a field
that, on the one hand, is not registered by the measuring devices,
and, on the other hand, is a component of the phase field.
Therefore, an analyzer may become the bifurcation domain for the
phase field.

An analyzer is a classical object. Interaction of the phase field
with a classical object may be of two types. Under the interaction
of the first type, the coherence of the phase field with the
emitting object remains intact; under the second, interaction
breaks the coherence. Since we assume that the phase field excites
oscillations in the analyzer that are coherent with the field, we
are to take that interaction with the analyzer does not break the
coherence of the field. We also take that the impact of the
quantum object on the analyzer is not registered macroscopically.
The registration tales place in the detector. The detector is a
classical system that is put into a state of unstable equilibrium.
The detector interacts with the kern of the quantum object. The
interaction results in bringing the detector out of equilibrium. A
catastrophic, macroscopically detectable process develops in the
detector. The detector (detectors) is placed at one (several)
output channels of the analyzer. Engaging of detector results in
fixation of the output channel of the analyzer that the kern of
the quantum object has hit.

In this way, the value of the observable $\A$ of the quantum
object becomes fixed. At the same time, an equivalence class that
includes the elementary state of the object under measurement
becomes fixed.

 The reverse impact of the detector onto the quantum
object is also strong. For an irreproducible measurement, the
elementary state of the object is changed completely. For a
reproducible measurement, coherence is broken anyway, but the
elementary state of the quantum object remains in the equivalence
class corresponding to the output channel that passed through
itself the kern of the quantum object. The kern of the quantum
object and other components of the phase field moving along with
the kern and passed through other output channels of the analyzer
cease to be coherent parts of phase field.

If a detector is placed on an output channel that did not convey
the kernel through, the detector experiences only a weak impact
from the part of the phase field that passed though the channel.
The catastrophic process does not develop in the detector, and no
macroscopic effect is registered. Despite this, feedback from the
detector onto the phase field is substantial. The field in this
channel loses coherence with the kernel of the quantum object, and
with the parts of the phase field that have passed through other
channels.

If detectors are not mounted in any of the output channels, it is
in principle possible to unite back all the parts of the phase
field transmitted through different channels. They will coherently
add up, and the initial elementary state may be reconstructed. If
there is a detector in one of the channels, the corresponding part
of the phase field cannot take part in the coherent addition.
Effectively, from the standpoint of elementary state of the
quantum object, this part of the phase field is lost.

Thus, a part of the phase field that determines the elementary
state of the quantum object may effectively "disappear" in two
cases. Either a change of the state (with decoherence) of the
disappeared part of the field takes place, or the disappeared part
does not change, but the state of the kernel of the quantum object
changes. In both cases, there is a change in the structure of the
phase field, which is coherent with the kernel. Namely, such a
field determines the elementary state of the quantum object. Under
a change of elementary state, a change of quantum state naturally
occurs. This change has all the features of the collapse of
quantum state induced by measurement. A similar model of
measurements is given in review~\cc{nam}.

The phase field performs the functions usually ascribed to hidden
parameters. However, in contrast to the situation with hidden
parameters, we point out the way to construct a mathematical image
of this field. For this reason, mathematically, there is no
existence problem for this field. All the objections that are
usually put forward against hidden parameters are not valid in the
case of a phase field. However, surely, the existence of material
realization of the phase field remains a hypothesis at the moment.

We point out that the status of this hypothesis differs from the
one of the statements from the above postulates. The latter
abstract the results obtained in physical observations. Based on
these postulates, the mathematical formalism of our scheme is
constructed. For this construction, the hypothesis we discuss is
not needed. On the contrary, within the classical paradigm, the
hypothesis can be derived from the mathematical formalism
constructed. More accurately, the phase field seems to be a most
viable candidate for the material realization of the mathematical
notion of elementary state. At the same time, the most viable does
not mean the only possible.

In contrast to hidden parameters, the phase field is partially
observable. It influences the kernel behavior of a quantum object,
if the kernel is coherent with the phase field. In turn, detector
of the classical measuring device is sensitive to the kernel. A
classical device is not sensitive in any way to a phase field that
lost coherence with its kernel. However, this does not mean that
this field has disappeared. It may manifest itself as dark matter.

The concept that any quantum object consists of a kernel and a
phase field allows for very intuitive interpretation of the
so-called delayed-choice experiment. The idea of this experiment
was suggested by Wheeler~\cc{wheel}; the experiment was performed
independently by two collaborations~\cc{all,hell}.

The scheme of the experiment is as follows. Figs. l(a) and l(b)
show two experimental setups. $M_1$ and $M_4$ are two
semireflective mirrors, $M_2$ and $M_3$ are normal mirrors, $D_A$
and $D_B$ are detectors.
\begin{center}
\begin{picture}(110,80)
\put(5,30){\vector(1,0){15}} \put(5,60){\vector(1,0){15}}
\put(20,30){\vector(1,0){25}} \put(5,65){\vector(0,-1){20}}
\put(5,45){\line(0,-1){15}} \put(35,60){\vector(0,-1){15}}
\put(5,70){\vector(0,-1){5}} \put(35,45){\vector(0,-1){25}}
 \put(20,60){\line(1,0){15}}
\put(35,20){\oval(5,7)} \put(45,30){\oval(7,5)}
\put(5,60){\circle{6}} \put(7,55){$M_1$} \put(27,55){$M_2$}
\put(7,33){$M_3$} \put(25,20){$D_B$} \put(43,35){$D_A$}
 \put(17,62){$B\rightarrow$}  \put(7,45){$A\downarrow$}
 \put(17,10){$(a)$}

 \put(60,30){\vector(1,0){15}} \put(60,60){\vector(1,0){15}}
\put(75,30){\vector(1,0){25}} \put(60,65){\vector(0,-1){20}}
\put(60,45){\line(0,-1){15}} \put(90,60){\vector(0,-1){15}}
\put(60,70){\vector(0,-1){5}} \put(90,45){\vector(0,-1){25}}
 \put(75,60){\line(1,0){15}}
\put(90,20){\oval(5,7)} \put(100,30){\oval(7,5)}
\put(60,60){\circle{6}} \put(90,30){\circle{6}} \put(62,55){$M_1$}
\put(82,55){$M_2$} \put(62,33){$M_3$} \put(82,33){$M_4$}
\put(80,20){$D_B$} \put(98,35){$D_A$}
 \put(72,62){$B\rightarrow$}  \put(62,45){$A\downarrow$}
 \put(72,10){(b)}
 \thicklines
\put(3.1,32){\line(1,-1){4}} \put(3.1,61.9){\line(1,-1){4}}
 \put(33,61.9){\line(1,-1){4}}

\put(57.9,32){\line(1,-1){4}} \put(57.9,61.9){\line(1,-1){4}}
\put(88,32){\line(1,-1){4}} \put(88,61.9){\line(1,-1){4}}
\end{picture}

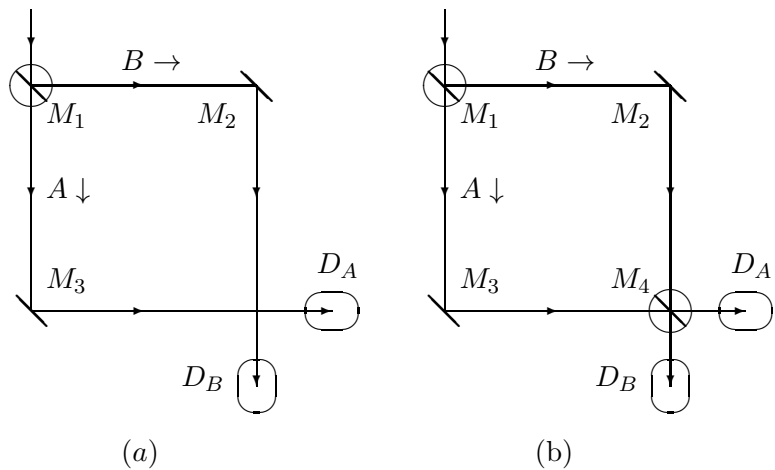
Fig. 1. Delayed-choice experiment.
\end{center}

A pulse of light arrives onto semireflective mirror $M_1$. The
pulse splits into two parts. One of the parts traverses path $A$,
another, path $B$. In the case of Fig. la, these parts do not mix
with one another and are registered independently by detectors
$D_A$ and $D_B$. In the case of Fig. lb, both parts get to the
semireflective mirror $M_4$, where interference between the two
parts takes place. Since the phase of oscillations changes by
$\pi/2$ at reflection, and does not change in passing through a
semireflective mirror $M_4$, all of the light pulse arrives at the
detector $D_B$ after the interference.

Wheeler's idea consists in making the pulse weak and short, and
the intervals between pulses large enough. In this way, not more
than a single photon is localized in the detector at any moment.
Second, the semireflective mirror $M_4$ is removable. By the
choice of the experimenter, it is either set or removed in the
time span between the moments when the photon passes the
semireflective mirror $M_1$ and reaches the domain of the
semireflective mirror $M_4$.

If the semireflective mirror $M_4$ is not put in (Fig. la), each
of the detectors $D_A$ and $D_B$ are engaged with probability 1/2
for each pulse. This case is easy to interpret if we consider the
photon as a particle. Then it takes either path $A$ or path $B$
after passing semireflective mirror $M_1$ with probabilities 1/2.
Correspondingly, it hits either detector $D_A$ or detector $D_B$.

If the semireflective mirror $M_4$ is put in (Fig. lb), only the
detector~$D_B$ responds. This case is easy to interpret if the
photon is considered as a wave. Then the wave is split by
semireflective mirror $M_1$ into two components propagating via
the paths $A$ and $B$. The components interfere at the
semireflective mirror $M_4$. After the interference, the wave
propagates after semireflective mirror $M_4$ only to the
detector~$D_B$.

The results of the experiments confirmed Wheeler's expectations
completely. This means that the photon behaves as a particle or as
a wave, depending on the situation. The zest of Wheelers idea is
that when a photon interacts with the semireflective mirror $M_1$;
it cannot know how it should behave-as a particle or as a wave. To
decide on this, it should foresee the whim of the experimentalist.

An orthodox partisan of standard quantum mechanics will say right
away that the above intuitive interpretations are wrong, because
they are classical. In reality, a photon is not a particle and not
a wave. After passing semireflective mirror $M_1$, neither a
photon nor its components propagate either along path $A$ or along
path $B$. What takes place in semireflective mirror $M_1$ is
really the spitting of the wave function of the initial photon
into parts corresponding to the paths $A$ and $B$. Next, in
semireflective mirror $M_4$, if it is present, the coherent
addition of these parts of the wave function takes place. Such
addition yields a correct account of the experiment.

Such an explanation is acceptable if one takes that wave function
describes a physical field. However, standard quantum mechanics
rejects this assumption. And it does so very justly, because, if
nothing else, the wave function takes values in complex numbers.
The wave function is probability amplitude. It is a purely
mathematical construct. As such, it cannot interact either with
semireflective mirror $M_1$ or with semireflective mirror $M_4$.
The evolution of the wave function should describe a physical
process. Standard quantum mechanics is unable to explain the
nature of this process.

The above experiment seems to confirm one of the major precepts of
quantum mechanics: a quantum object cannot have any definite
trajectory. Despite this precept, almost all experiments in
particle physics (which are surely quantum objects) are based on
analysis of these non-existing trajectories. The situation is
paradoxical: theoreticians and experimenters in elementary
particle physics play simultaneously the same game obeying
different rules.

The hypothesis of the phase field supports the experimenter's
interpretation. After the semireflective mirror $M_1$; the
photon's kernel propagates either by path $A$ or by path $B$
depending on its elementary state. The phase field splits into two
components on semireflective mirror $M_1$ the components propagate
by different paths. In the mirror $M_4$, if it is present, these
two components of the phase field add up coherently into the
resulting phase field (elementary state) that directs the photon's
kernel to the detector $D_B$.

Also, within our hypothesis, the scattering of a quantum object
off two slits can be interpreted. In the experiment, a distinct
interference pattern is observed. The interference pattern is
observed even in the case when the intensity of the particle beam
is so weak that not more than one particle is located in the
experimental setup~\cc{ton}. In this case, it is not possible to
explain the interference by interaction between incident
particles. Rejecting verbal embellishments, the standard
interpretation of this experiment reduces to the following.
Indivisible before the slits, the quantum object passes
simultaneously through both slits, which are separated by a
macroscopic distance; after that, the object stays indivisible.

Our interpretation is much more intuitive. The phase field of the
quantum object excites weak collective oscillations in the screen.
These oscillations generate a secondary classical field.
Oscillations in various parts of the screen are coherent with one
another. Because of this, the components of the secondary field
radiated by different parts of the screen add up coherently. The
kernel of the quantum object approaches as a whole the slits. The
kernel may scatter off the slit at different angles. From the
standpoint of standard quantum mechanics, this process has no
definite cause. In terms of this paper, the slit is a bifurcation
domain for the quantum object. The behavior of a particular
quantum object in such a domain is governed by a random force.
This random force results from interaction between the kernel and
the secondary classical field radiated by the screen. This field
is very weak, but it is coherent with the kernel. Because of this,
its interaction with the kernel is a resonance one. The structure
of the resonance field at the location of the kernel changes when
one of the slits is closed. In view of this, the ensemble of
quantum objects scattered off two slits is not a simple mix of
ensembles of quantum objects scattered separately by each of the
slits. As a result, the interference pattern appears.

Surely, these qualitative considerations are insufficient for
quantitative computation of the interference pattern. For such a
computation, mathematical formalism of the standard quantum
mechanics can be used. This is the case because the interference
pattern appears only if a very large number of quantum objects is
scattered (see~\cc{ton}). In this case, we deal not with a single
elementary state, but with a large number of elementary states. By
the law of large numbers, this collection of elementary states can
be replaced with an equivalence class, i.e., with a quantum state.
Here, we get to the domain of applicability of the standard
quantum mechanics.

\section{CONCLUSIONS}

The approach we have presented by no means rejects standard
quantum mechanics. The original researches did a splendid job in
developing quantum mechanics, but they started the construction
from the first floor-from the description of probabilities and
average values. Because of this, the stability of the construction
required the introduction of a number of props in the form of
"principles": the superposition principle, the uncertainty
principle, the complementarity principle, the projection
principle, the indistinguishability principle, and the no
trajectories principle. All of these are far from obvious, and
appear to be rather artificial. On the other hand, a number of
principles-causality, formal logic-had to be rejected, despite the
fact that they had stood the test of all of the previous history
of mankind. It may seem that these two are only deeply rooted
misconceptions. Examples of such misconceptions are available. On
the other hand, we should do our best to keep them alive. The
scheme we advocate is an attempt in this direction.

In a way, the relation of our scheme to the standard quantum
mechanics resembles the one of statistical physics to
thermodynamics. Thermodynamics can be constructed on its own,
without reference to statistical physics. Historically, this
indeed took place. It is also true, however, that thermodynamics
gained extra momentum after its basic principles had lost the
status of fundamental laws of nature. Instead, these principles
became derivable from the more fundamental laws of statistical
physics. In this case, physics gained the possibility to use to
the full extent the powerful formalism of probability theory.

A distinctive feature of statistical physics is that it involves
the notion of elementary event. In statistical physics, which
studies physical systems consisting of a huge number of
components, an elementary event is the state of all the
components. In practice, one cannot observe or fix it. However,
its existence is a matter of principle, because it allows one to
use probability theory.

In standard quantum mechanics, as it was in thermodynamics, the
notion of elementary event is absent. The principal feature of our
approach is that it introduces the notion of elementary state,
which can be considered as an elementary event. As in statistical
physics, an elementary state is unobservable, but allows one to
use to the full extent the classical formal logic and the standard
probability theory.

Using the notion of elementary state, it is possible to establish
the limits of applicability of the mathematical formalism of
standard quantum mechanics. Within these limits is the study of
the ensembles of quantum objects that can be described by a
quantum state. It is a very important type of ensembles, but it is
far from being the most general type. In particular, standard
quantum mechanics is unable to describe individual quantum
objects. At the same time, the formalism of modern probability
theory is powerful enough to describe the behavior of ensembles of
a more general type. Thus, we hope that the range of applicability
of quantum mechanics may be extended.

Moreover, with an elementary state that corresponds to individual
quantum object, it is possible to give an intuitive interpretation
of quantum phenomena. The notion of elementary state allows one to
get rid of so-called quantum logic and quantum probability theory.
Both schemes are presently more advertisements than developed
schemes. Intuitively, we perceive them as unfeasible.

The scheme we have suggested makes it possible to give a unified
treatment of quantum and classical systems. Due to this, a
question rather painful for the standard quantum mechanics-which
mechanics, quantum or classical, is logically primary-is resolved.

Finally, our scheme does not reject the causality principle,
treating it as a principle shared by both quantum and classical
physics.

\end{document}